%% file: paper.tex
\documentclass[sigplan,10pt]{acmart}

\renewcommand\footnotetextcopyrightpermission[1]{}  
\setcopyright{none}
\AtBeginDocument{
  }
    
\usepackage{algorithmic}
\usepackage{graphicx}
\usepackage{textcomp}
\usepackage{array}
\usepackage{xcolor}
\usepackage{subcaption}
\usepackage{booktabs}
\usepackage{multirow}
\usepackage{listings}
\usepackage{adjustbox}
\usepackage{tabularx}
\usepackage{booktabs}
\usepackage[ruled,vlined]{algorithm2e}
\usepackage{hyperref}
\usepackage{cleveref}

\newcommand{\Mark}[1]{ {\color{teal}{Mark:#1}} }

\newcommand{\ignore}[1]{}

\newcommand{\name}{VOLT}

\begin{document}

%\title{\name{}: Extensible Compiler Framework for Open GPUs}
%\title{Challenges of designing open source GPU compiler}
\title{Inside VOLT: Designing an Open-Source GPU Compiler}

\author{Shinnung Jeong}
\email{sjeong306@gatech.edu}
\affiliation{%
  \institution{Georgia Tech}
  \country{USA}
}

\author{Chihyo Ahn}
\email{ahnch@gatech.edu}
\affiliation{%
  \institution{Georgia Tech}
  \country{USA}
}

\author{Huanzhi Pu }
\email{hpu8@gatech.edu}
\affiliation{%
  \institution{Georgia Tech}
  \country{USA}
}

\author{Jisheng Zhao}
\email{jishengz@gmail.com}
\affiliation{%
  \institution{Georgia Tech}
  \country{USA}
}

\author{Hyesoon Kim }
\email{hyesoon@cc.gatech.edu}
\affiliation{%
  \institution{Georgia Tech}
  \country{USA}
}

\author{Blaise Tine}
\email{blaisetine@cs.ucla.edu}
\affiliation{%
  \institution{University of California, Los Angeles}
  \country{USA}
}

\crefformat{section}{\S#2#1#3}
% \crefname{lstlisting}{Listing}{Listings}
% \Crefname{lstlisting}{Listing}{Listings}
\crefname{section}{Section}{Sections}
\crefname{figure}{Figure}{Figures}
\crefname{table}{Table}{Tables}
\crefname{alg}{Algorithm}{Algorithms}
\crefname{equation}{Equation}{Equations}
\creflabelformat{equation}{#2\textup{#1}#3}
\crefname{appendix}{Appendix}{Appendices}
\crefname{algorithm}{Algorithm}{Algorithms}
\crefname{lstnumber}{line}{lines}
\crefname{lstnumber}{Line}{Lines}
\crefname{lstlisting}{listing}{listings}
\Crefname{lstlisting}{Listing}{Listings}
\crefname{listing}{Listing}{Listings}
\Crefname{listing}{Listing}{Listings}
\renewcommand{\crefrangeconjunction}{--}

\input{paper/0.abstract}


\maketitle

\input{paper/1.introduction}

\input{paper/2.background}
\input{paper/3.0.motivation}
\input{paper/3.1.Design}
\input{paper/3.2.Eval}
\input{paper/6.case_study_arch}

\input{paper/5.case_study_frontend}

\input{paper/7.related_works}
\input{paper/8.discussion}
\input{paper/9.conclusion}

\bibliographystyle{ACM-Reference-Format}
\bibliography{paper}

\appendix

\end{document}

%% file: paper/0.abstract.tex
\begin{abstract}

Recent efforts in open-source GPU research are opening new avenues in a domain that has long been tightly coupled with a few commercial vendors. Emerging open GPU architectures define SIMT functionality through their own ISAs, but executing existing GPU programs and optimizing performance on these ISAs relies on a compiler framework that is technically complex and often undercounted in open-hardware development costs. 

To address this challenge,  the Vortex-Optimized Lightweight Toolchain (\name{}) has been proposed. This paper presents its design principles, overall structure, and the key compiler transformations required to support SIMT execution on Vortex.
\name{} enables SIMT code generation and optimization across
multiple levels of abstraction through a hierarchical design that accommodates diverse front-end languages and open GPU hardware. To ensure extensibility as GPU architectures evolve, \name{} centralizes fundamental SIMT-related analyses and optimizations in the middle-end, allowing them to be reused across front-ends and easily adapted to emerging open-GPU variants. Through two case studies on ISA extensions and host-runtime API, this paper also demonstrates how \name{} can support extensions.

%Through two case studies covering ISA extensions and host–runtime memory API extensions, we demonstrate how \name{}’s modular design supports concrete extensions across both compiler back-end and runtime layers.

%To address this challenge, we introduce a new end-to-end compiler framework for open GPU architectures, focused on supporting the Vortex open-source GPU, named the Vortex-Optimized Lightweight Toolchain (\name{}). 
\end{abstract}

%% file: paper/1.introduction.tex
\section{Introduction}

%related works~\cite{cc:14:taming, arxiv:24:cf_management, caches:11:characterization, cgo:22:dram, isca:12:capri, hpca:13:dual-path, hpca:14:multipath, hpcs:16:cf_restructuring, ics:15:dacache, micro:14:exploringDSE, pact:11:divergence, pact:12:data_herding, popl:03:folklore}

In modern computing, GPUs have become indispensable hardware across a wide range of domains, including artificial intelligence and high-performance computing~\cite{paper:survey:gpu, paper:survey:parrallelcomputing:14, paper:survey:ai_trends}. GPUs enable massive parallelism and high-throughput computation, unlocking substantial performance gains for data-intensive and parallelizable applications. As their role continues to grow in both scientific and industrial workloads, understanding and controlling the full GPU software stack has become increasingly important.

However, GPU research and development remain tightly coupled to a few commercial vendors such as NVIDIA, AMD, or Intel. Most of the underlying architecture, runtime behavior, and compiler toolchains are proprietary and closed, limiting community insight and innovation through realistic hardware. Recently, several efforts have emerged to develop open-source GPUs~\cite{vortex, vortex_first_2020, iccd:24:ventus, arxiv:25:egpu, fpga:16:fgpu, fpl:21:ma, fpl:12:guppy, micro:17:scrach, vortex:skybox, vortex:ipdpsw, vortex:sparseweaver, dateworkshop:25:pu, dateworkshop:25:Bonet, asplos:25:vringo, carrv:simty, rv64x, asplos:25:vringo,iccd:24:ventus, iccd:24:SIMTight}, with the goal of bringing GPU hardware into the domain of transparent and reproducible research, enabling broader participation and experimentation.

Among recent efforts in open-source GPU research, one of the most active and distinctive projects is the Vortex GPU~\cite{vortex, vortex_first_2020}, an open-source RISC-V-based GPU that achieves general-purpose GPU execution with additional instructions for thread management and graphics. 
However, while the hardware demonstrates feasibility, fully realizing its potential in real-world programs remains a challenge. In particular, effective binary translation, optimized code generation, and executing existing programs using diverse frontend languages remain challenging, revealing a gap between ISAs and programs.

This highlights a key gap: while open-source GPU hardware is rapidly advancing, \textit{compiler} support remains comparatively underdeveloped—despite being a specialized, non-trivial engineering effort that is often undercounted in open-hardware project plans. Commercial vendor–provided compilers are typically closed-source, even though a substantial portion of GPU performance depends on compiler behavior, from scheduling and register allocation to low-level code generation. Their proprietary nature limits transparency and portability. The lack of visibility into these back-ends hinders the diagnosis of performance bottlenecks and code generation issues, ultimately slowing architectural exploration.

\ignore{
\begin{figure}[t]  
    \centering
    \includegraphics[width=0.48\textwidth]{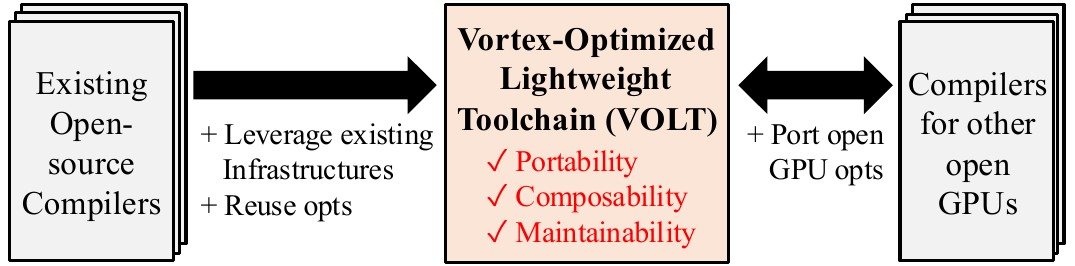}
    \caption{\name{} framework for open GPUs}
    \label{fig:intro}
    \vspace*{-\textfloatsep}
\end{figure}
}

\begin{table*}[t]
\centering
\footnotesize
\caption{Summary of VOLT’s contributions across the toolchain. We count line of code(LoC) changes relative to upstream. “X” means non-intrusive. }
\vspace{-2mm}
\begin{tabular}{p{2.2cm} p{2.2cm} p{3.8cm} p{5cm} p{1.3cm} p{0.9cm}}
\toprule
\textbf{Stage} & \textbf{Baseline} & \textbf{Prior Limitations} &
\textbf{VOLT Contribution} & \textbf{Intrusive?} & \textbf{LoC} \\
\midrule
OpenCL Front-end & PoCL~\cite{paper:pocl:2015} &
No existing support for the Vortex GPU &
Kernel-function translation, host-API translation, memory model integration, and scheduling support for the Vortex GPU &
X & 1565 \\
\midrule
CUDA Front-end & CuPBoP~\cite{fccm:25:softcuda} &
Lack of host runtime API and kernel function support &
Extended host and kernel function support for Vortex GPU (see case studies) &
X & 4155 \\
\midrule
Middle-End & LLVM~\cite{paper:llvm} &
Absence of SIMT-style uniformity and divergence analysis applicable to open-source GPU targets&
Divergence analysis, transformation and optimization passes, and divergence-management intrinsic insertion&
X & 4653 \\
\midrule
Back-end & LLVM RISC-V~\cite{paper:llvm} &
Absence of Vortex ISA support &
Extension of the ISA table and target code generation &
X & Included above \\
\midrule
End-to-End & LLVM + PoCL + CuPBoP &
No existing CUDA/OpenCL compilation pipeline for the Vortex &
First fully open-source CUDA/OpenCL compilation pipeline targeting the Vortex GPU &
X & -- \\
\bottomrule
\end{tabular}

\label{tab:volt-contrib}
\vspace{-2mm}
\end{table*}

These challenges lead to the development of a fully open-source compiler infrastructure within the Vortex GPU project, called the Vortex-Optimized Lightweight Toolchain (\name{}). At the core of \name{} lies \textit{extensibility}, which enables IR-level optimizations and analyses to be reused across heterogeneous GPU ISAs. This extensibility not only strengthens openness but also alleviates the engineering burden of maintaining separate compiler stacks for each Vortex GPU variant or future generation.

This paper provides the first tool-oriented description of \name{}, presenting its design challenges, design principles, and complete end-to-end structure.
A central design challenge behind \name{} is determining how to maximize portability while supporting a rapidly evolving open GPU architecture. This challenge divides into two  questions:
(1) how to implement and share SIMT-aware analyses and optimizations in a way that remains portable across Vortex variants and potentially other open GPUs, and (2) how to support multiple front-end languages (particularly two important GPU language, OpenCL and CUDA) without creating a maintenance burden.

For the first challenge, the key insight is to centralize fundamental SIMT-aware analyses and transformations in the middle-end, enabling architecture-agnostic behavior that remains stable as the ISA evolves.
Vendor toolchains and other open GPUs~\cite{iccd:24:ventus} typically implement divergence handling only at the machine-IR level, which limits cross-architecture reuse and increases engineering burden, especially for open-source projects with limited manpower.
By contrast, \name{} lifts these core SIMT semantics to the IR level, forming a portable and analyzable foundation for GPU compilation.

The second challenge is addressed through explicit attention to composability and maintainability.
\name{} adopts a hierarchical organization spanning front-end language semantics, middle-end IR transformations, and back-end code generation, while building on composable compiler-infrastructure components rather than reimplementing entire stacks.
This design reduces engineering cost and enables rapid experimentation with ISA extensions, analyses, and runtime semantics. As summarized in \cref{tab:volt-contrib}, \name{} extends existing infrastructures~\cite{paper:pocl:2015, paper:llvm} and integrates external tools~\cite{fccm:25:softcuda} to support CUDA.

\name{} further supports new ISA features and runtime behaviors by extending modular components rather than modifying the core pipeline.
We show this extensibility through two case studies: one showing how \name{} integrates evolving GPU ISA extensions, and another demonstrating how it adds host–runtime API support, including CUDA memory APIs adapted to the Vortex memory model.

This paper makes the following key contributions:
\begin{itemize}
    \item Presents the design principles and hierarchical compiler design of \name{}.

    \item Describes the centralized and portable, SIMT-aware transformations implemented in \name{} in detail.
    
    \item Show the extensibility of \name{} along two primary axes: ISA extensions and host–runtime extensions.

\end{itemize}

%% file: paper/2.background.tex
\section{Background and Related Works}

\subsection{Existing Open GPUs and their ISA}
\label{sec:back:gpus}

In recent years, multiple efforts have sought to open up GPU design, resulting in fully open-source GPU implementations, 
based on RISC-V ISA~\cite{iccd:24:SIMTight, VeriGPU}, RISC-V with Vector extensions~\cite{iccd:24:ventus, carrv:simty}, RISC-V with Vortex ISA extension~\cite{vortex, vortex_first_2020, vortex:skybox, vortex:sparseweaver, asplos:25:vringo, arxiv:25:egpu, dateworkshop:25:Bonet, dateworkshop:25:pu}.
These existing open GPUs highlight two trends.
First, many open GPU projects are based on the RISC-V ISA. Since RISC-V supports an extensible ISA and architecture, open-source GPUs often extend and adopt the RISC-V architecture and ISA~\cite{manual:riscv}.
Second, the Vortex and its ISA have been adopted by several recent open-source GPU projects~\cite{vortex:skybox, vortex:sparseweaver, asplos:25:vringo, arxiv:25:egpu, dateworkshop:25:Bonet}.

\subsection{Existing Compiler Supports for Open GPUs}

To the best of our knowledge, non-Vortex open GPU projects likewise do not published papers for a full compiler stack, nor do they emphasize portability as a core design goal.
Ventus~\cite{iccd:24:ventus} uses LLVM~\cite{paper:llvm} and PoCL~\cite{paper:pocl:2015} for compilation, but its optimization passes, including key SIMT-aware transformations such as divergence management, are implemented in the back-end.
Other projects~\cite{rv64x, carrv:simty, fpga:16:fgpu} rely on LLVM for RISC-V code generation and introduce microarchitectural mechanisms for SIMT execution (e.g., via the RISC-V Vector ISA), while others use GCC~\cite{iccd:24:SIMTight} or HIP~\cite{VeriGPU}.
However, these efforts describe only partial components rather than end-to-end compiler frameworks designed with portability and long-term extensibility in mind.

Likewise, no prior publication has described a complete end-to-end compiler framework for the Vortex GPU. Existing Vortex-related work focuses on individual parts of the software stack: the architecture and ISA~\cite{vortex}, 3D graphics support and limited discussion of divergence management~\cite{vortex:skybox}, and FPGA-based hardware evaluations using CUDA~\cite{fccm:25:softcuda}. None of these works provides a detailed description of a full compilation pipeline, nor do they explain detailed the SIMT-aware analyses, transformations, or optimizations required to generate Vortex GPU binary.

\subsection{ISA Extension Requirements for Open GPUs}

Although many open-source GPU projects adopt RISC-V, its CPU-oriented design necessitates additional ISA mechanisms to properly express GPU-specific execution semantics. While RISC-V already provides a Vector Extension for SIMD(Single Instruction, Multiple Data) execution that includes vector operations, data movement, and type conversions for vector processors, these features alone are insufficient to support SIMT(Single Instruction, Multiple Threads)-based GPU execution.

\begin{figure}[t]
    \centering
    \begin{subfigure}[b]{0.13\textwidth}
        \centering
        \includegraphics[width=\textwidth]{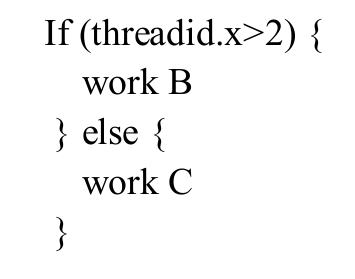}
        \caption{Example Code}
        \label{fig:moti:intro:code}
    \end{subfigure}%
    \hfill
    \begin{subfigure}[b]{0.13\textwidth}
        \centering
        \includegraphics[width=\textwidth]{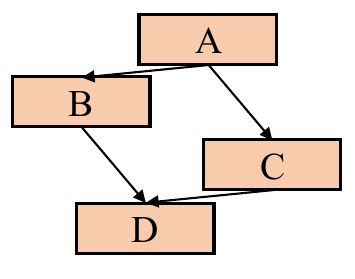}
        \caption{CFG}
        \label{fig:moti:intro:cfg}
    \end{subfigure}%
    \hfill
    \begin{subfigure}[b]{0.12\textwidth}
        \centering
        \includegraphics[width=\textwidth]{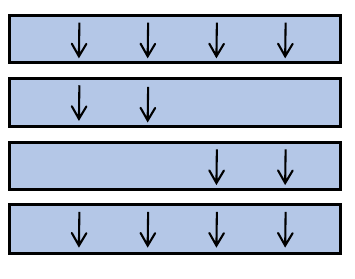}
        \caption{Thread Execution}
        \label{fig:moti:intro:thread}
    \end{subfigure}
    \caption{Illustration of thread execution with control-flow divergence (assuming a total of four threads).}
    \label{fig:moti:graph}
    \vspace{-5mm}
\end{figure}

The distinction between SIMD and SIMT is crucial for compiler design. SIMD operates with a single instruction pointer (PC) executing over multiple data elements, originally introduced as special-purpose CPU instructions. By contrast, SIMT introduces the concept of wavefronts (small thread groups executing in lockstep). Each wavefront maintains its own PC and stack, and multiple wavefronts can execute concurrently with divergent control flow, thereby enabling parallelism even when execution paths differ. Furthermore, in SIMD, if some elements follow a different branch, the divergent control flow must still be serialized within a single thread context. However, SIMT allows distinct wavefronts to maintain independent PCs to handle such divergence. Within a single wavefront, only the threads that must execute the branch are marked via a thread mask and executed accordingly, as shown in \cref{fig:moti:graph}.

In addition, the hardware microarchitecture itself manages divergence-related information such as the Immediate Post-Dominator (IPDOM) stack, ensuring correct divergence and reconvergence of threads.
For correct execution, the IPDOM stack requires structured control flow: each divergence point must correspond to a matching reconvergence point, and that reconvergence point must be the immediate post-dominator of the split.
To meet this requirement, the compiler must ensure that the CFG reflects these well-nested regions so that the IPDOM stack can correctly push and pop divergence states without misaligned reconvergence.

Therefore, GPU ISA needs to support the following core operations, which are not considered for CPU and SIMD execution. First, the ISA must specify the spawn points of wavefronts to indicate where GPU-style execution begins, including information such as the number of wavefronts to be launched and the initial program counter (PC) address. Second, it must provide instructions to set or retrieve the thread mask, representing which threads are active at a given point in execution. Third, the ISA must include mechanisms to explicitly represent control-flow branches and their reconvergence points, passing this information to the hardware responsible for divergence management. To fully capture the notion of wavefront execution and to manage divergence, additional GPU-oriented ISA extensions are required.

\subsection{Vortex ISA for GPU Core Operations}
\label{sec:back:vortex_isa}

\begin{table}[t]
\centering
\footnotesize
\caption{Vortex ISA Extensions~\cite{vortex}}
\vspace{-2mm}
\begin{tabular}{>{\raggedright\arraybackslash}p{3cm}|p{5cm}}
\hline
\textbf{ISA Extension} & \textbf{Description} \\
\hline\hline
\textbf{vx\_wspawn} \texttt{\#warps, \%PC}  & Spawns \texttt{\#warps} at program counter \texttt{\%PC} \\ \hline
\textbf{vx\_tmc} \texttt{tmask}    & Sets the active threads for the warp using \texttt{tmask} \\ \hline
\textbf{vx\_active\_threads} \texttt{tmask} & Return active thread mask\\ \hline
\texttt{\#ipdom\_addr} \texttt{<-} \textbf{vx\_split} \texttt{\#pred}  & Marks the beginning of a divergent branch, push the current state into the IPDOM stack, and returns its stack address \\ \hline
\textbf{vx\_join} \texttt{\#ipdom\_addr}   & Re-converges threads at the post-dominator and pops the IPDOM stack \\ \hline
\textbf{vx\_pred} \texttt{\#pred, tmask}   & Evaluates the loop predicate and retrieves the \texttt{tmask} after completing loop iterations \\ \hline
\textbf{vx\_barrier} \texttt{\#id, (\#warps or \#cores)} & Synchronizes a wavefront of a local barrier or a global barrier \\ \hline
\end{tabular}
\label{tab:isa_instructions}
\vspace{-2mm}
\end{table}

\begin{figure}[t]
    \centering
    \begin{subfigure}[b]{0.19\textwidth}
        \centering
        \includegraphics[width=\textwidth]{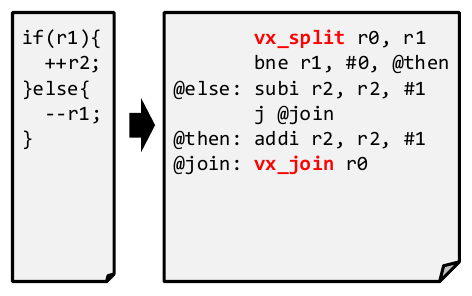}
        \caption{If-Else}
        \label{fig:design:ex:if}
    \end{subfigure}%
    \hfill
    \begin{subfigure}[b]{0.23\textwidth}
        \centering
        \includegraphics[width=\textwidth]{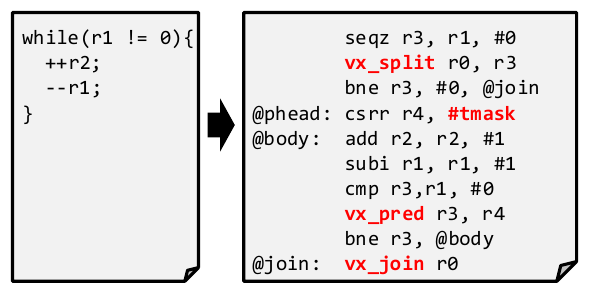}
        \caption{Loop}
        \label{fig:design:ex:loop}
    \end{subfigure}%
    \caption{Example of machine-level code for control-flow constructs: if-else and loop}
    \label{fig:design:ex}
\vspace{-5mm}
\end{figure}

To efficiently support GPU core operations, Vortex GPU~\cite{vortex_first_2020, vortex} suggests ISA extension as summarized in \cref{tab:isa_instructions}.
First, the \textbf{vx\_wspawn} instruction is introduced to spawn warps from a given address. 
In typical GPU programming models, the program grid is determined by runtime arguments (e.g., the number of blocks and threads specified in CUDA). Consequently, the number of wavefronts to be spawned varies during the runtime information. At the beginning of the initialization stage, a single thread runs to determine how many wavefronts and cores should be launched, while also generating the necessary metadata for subsequent execution. Once this initialization process completes, the program executes \textbf{vx\_wspawn} to spawn the threads.

Second, to configure and manage the active threads within a warp, Vortex introduces \textbf{vx\_tmc}, \textbf{vx\_active\_threads}, and \textbf{vx\_barrier} instructions. The \textbf{vx\_tmc} instruction takes a thread mask as an operand and activates the corresponding threads in the warp accordingly.
\textbf{vx\_active\_threads} returns the current active thread mask.
The \textbf{vx\_barrier} instruction specifies a wavefront-local or a global barrier.

Third, to support thread divergence of if–else and loops, Vortex introduces three additional instructions: \textbf{vx\_split}, \textbf{vx\_join}, and \textbf{vx\_pred}.
The \textbf{vx\_split} instruction marks the beginning of a divergence region, pushes the current thread mask and related metadata onto the IPDOM stack, \emph{and returns the stacked address of the pushed state}. The \textbf{vx\_join} instruction marks a reconvergence point, retrieving the saved state from the IPDOM stack to resume unified execution.
Finally, \textbf{vx\_pred} supports predicated loops: it evaluates a loop predicate and updates the thread mask to reflect which threads should continue executing the next iteration. Once no threads satisfy the loop condition, \textbf{vx\_pred} restores the original thread mask and jumps to the loop-exit block.

\Cref{fig:design:ex} illustrates how divergence-control instructions are used in machine-level code. \Cref{fig:design:ex:if} shows the transformation of an if–else branch, where \textbf{vx\_split} indicates the divergence point and \textbf{vx\_join} indicates the reconvergence point. \Cref{fig:design:ex:loop} presents the transformation of a while loop. Divergence control is handled by using \textbf{vx\_split} and \textbf{vx\_join} to evaluate the initial condition, while \textbf{vx\_pred} is used to evaluate the loop predicate in each iteration. In this process, only threads that satisfy the loop condition remain active. If none of the currently active threads satisfy the loop predicate, all threads that initially entered the loop are reactivated, and execution proceeds to the join block without taking the \texttt{bne} instruction, thereby terminating the loop.

\begin{figure}[t]
    \centering
    \includegraphics[width=0.48\textwidth]{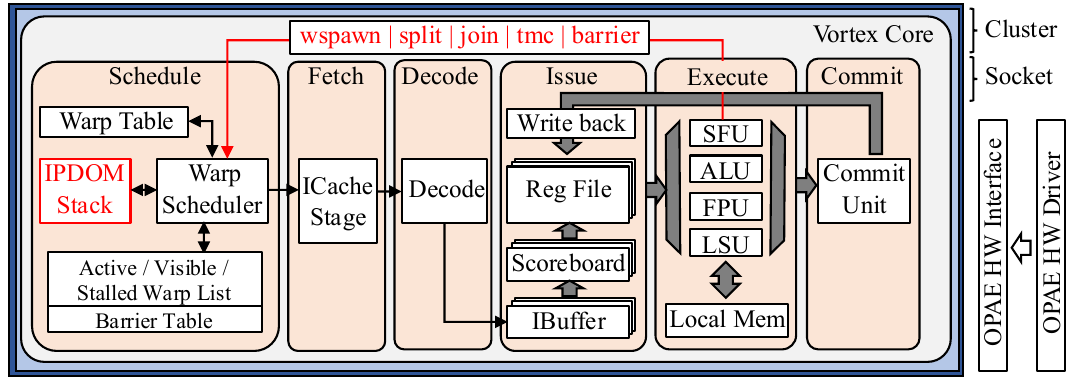}
    \caption{Vortex architecture~\cite{vortex}}
    \label{fig:vortex:arch}
    \vspace*{-\textfloatsep}
\end{figure}

\Cref{fig:vortex:arch} illustrates the Vortex GPU architecture with ISA extensions. To support these extensions, the pipeline incorporates a Special Function Unit (SFU) that executes Vortex-specific ISA instructions and updates the warp scheduler accordingly. In the \emph{schedule} stage, the microarchitecture maintains four key data structures closely related to divergence and synchronization management: (i) an \textit{IPDOM stack} per warp for tracking divergence and reconvergence state; (ii) a \textit{warp table} recording each warp's PC, active-thread mask; (iii) a \textit{barrier table} that tracks wavefront-local and global barriers; and (iv) \textit{active and stalled} warp lists that drive issue selection. Beyond these additions, the Vortex GPU retains a hierarchical organization consisting of clusters, sockets, cores, warps, and threads, and employs a six-stage pipeline with in-order issue and out-of-order commit. The architecture is further characterized by high reconfigurability, enabling customization of nearly all aspects of the hierarchy, including the number of units, the size of core execution units, and the memory and cache capacities. Owing to this flexibility and simple ISA extension, several recent works~\cite{vortex:skybox, vortex:ipdpsw, vortex:sparseweaver, arxiv:24:cf_management, arxiv:25:egpu, dateworkshop:25:Bonet, dateworkshop:25:pu, fccm:25:softcuda, asplos:25:vringo, hpca:25:tensor_core, hpca:25:lmi} have extended and leveraged the Vortex GPU to implement and evaluate their own designs.

%% file: paper/3.0.motivation.tex
\section{Motivation and Design Principles}
% list of problems? 
% This is the first paper describing an open-source GPU compiler - challenges to develop it? Provide a full list of the major components that need to be added. challenges? 
% multiple frontend 
% compare with Ventus? 

\subsection{Challenges of Open Source Full-Stack Projects}
Even with open GPU ISAs such as the Vortex ISA, building practical compilers for open GPUs remains challenging. A GPU compiler must support diverse front-end languages and implement a full optimization pipeline that handles SIMT-aware features such as divergence control, uniformity analysis, and host–runtime interaction. Robust compiler infrastructure is therefore essential not only for correctness but also for enabling performance portability and broader adoption of open GPU research. Because many open GPU projects adopt RISC-V, achieving ISA portability is both practical and highly desirable. 

These challenges become even more pronounced in open-source project, where engineering resources are limited. A maintainable and modular design is thus crucial for long-term sustainability. \name{} addresses this need by leveraging LLVM as a unified middle-end and back-end framework and by targeting Vortex ISA extensions in the back end. This structure supports both base RISC-V and Vortex ISAs while naturally aligning with the portability and composability.

\subsection{Design Principles of \name{}}

The design of \name{} is guided by four core principles. 
While these principles are long established in traditional compiler design, they are also equally essential for GPU.

\textbf{Openness} refers to transparency that enables external understanding and extension.
\name{} regard openness as a fundamental goal in compiler design, as it promotes accessibility, reproducibility, and collaborative development.
By making \name{} observable, openness encourages community-driven evaluation and allows contributors to experiment with new optimizations or back-end targets without upstream access.

\textbf{Portability} encompasses (1) operability across diverse hardware and platforms and (2) the reuse of existing optimizations on other architectures.
To support a wide range of hardware targets, a compiler should extend to multiple ISAs with minimal modifications.
Therefore, \name{} centralize key analysis, transformation, and optimization in the middle-end compiler to exploit the reuse  optimizations.
This requires designing abstractions that decouple optimizations from back-end-specific behavior.

\textbf{Composability} refers to the ability to flexibly assemble compiler components through loose coupling and clearly defined interfaces.
In rapidly evolving GPU front-end language and hardware, this design is essential. \name{} adopt hierarchical compiler design and each compiler should expose clear input/output contracts, allowing independent development and reuse.
Such a structure minimizes inter-component dependencies and enables rapid prototyping without destabilizing the overall system.

%In rapidly evolving GPU environments, composable design is essential.
%To support this, each compiler pass should be defined with explicit input and output contracts, enabling independent development and reuse.
%Such a loosely coupled structure reduces dependencies between components, enabling rapid prototyping and iterative development without introducing system-wide disruptions.

\textbf{Maintainability} refers to how easily a compiler can be understood, modified, and extended over time.
It is essential for long-term sustainability, especially in open-source projects with distributed contributors.
To support maintainability, \name{} avoid designing the entire compiler infrastructure from scratch and instead reuse existing open-source components and compiler passes.

%% file: paper/3.1.Design.tex
\section{\name{} Compiler Framework}
\label{sec:framework}

\subsection{Framework Overview}

\begin{figure}[t]
    \centering
    \includegraphics[width=0.44\textwidth]{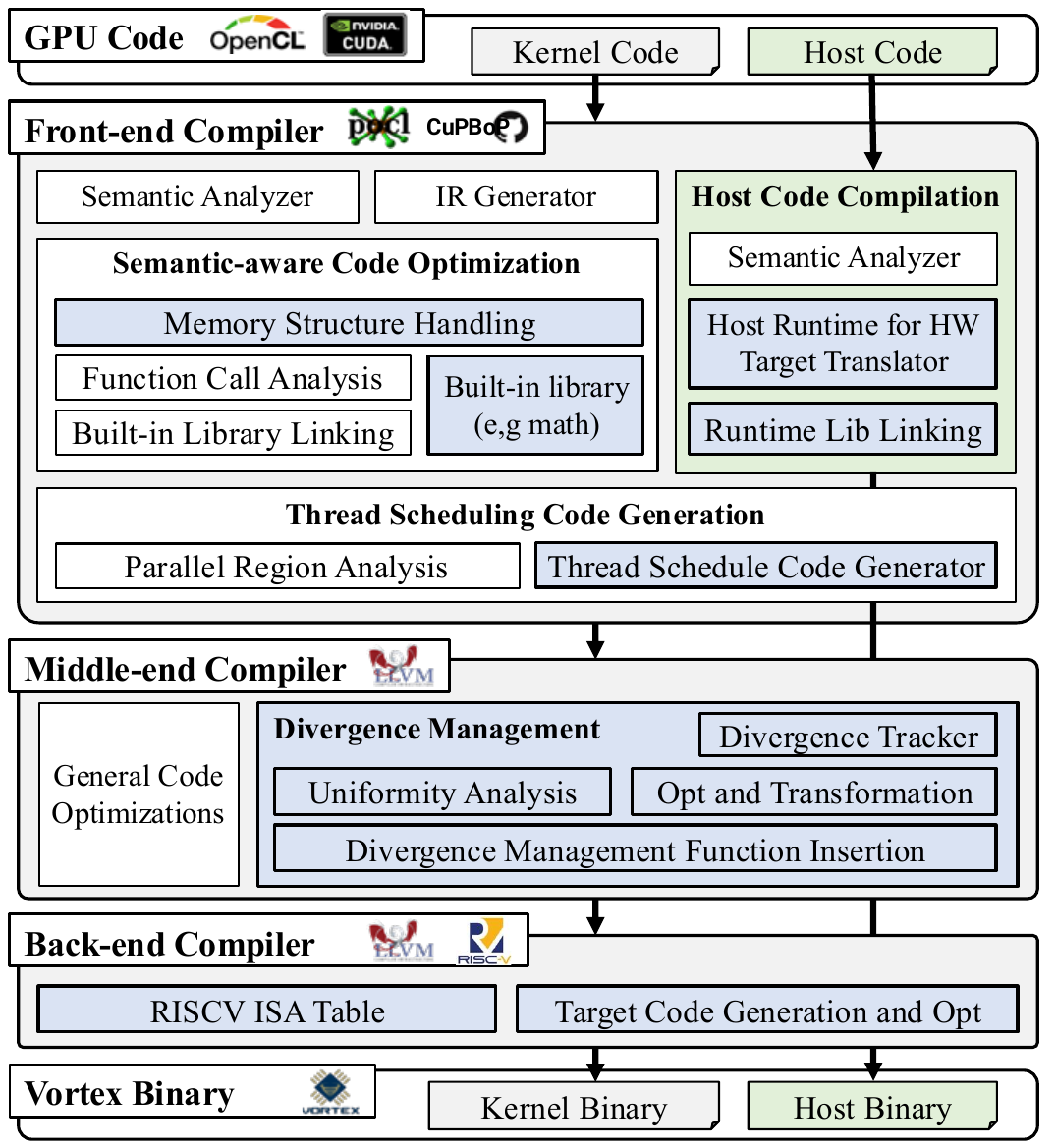}
    \caption{Overview of the \name{} framework. Blue components are proposed or extended to support the Vortex GPU. Green boxes represent the host-side compilation flow, and others represent the kernel compilation flow.}
    \label{fig:design:framework}
    \vspace{-5mm}
\end{figure}

As shown in \cref{tab:volt-contrib}, \name{} comprises a three-layered hierarchical design (front-end, middle-end, and back-end) built upon open-source projects. To manage increasing architectural diversity and to support multiple front-end languages with strong portability and composability, the compiler is structured into distinct layers, each responsible for a specific compilation stage. This layered design simplifies hardware-specific extensions and front-end integration while keeping complexity low through proper abstraction and separation of concerns. By building \name{} on existing open-source compiler infrastructures, \name{} also ensures long-term maintainability.

\lstset{
  float,                  
  captionpos=b,           
  basicstyle=\scriptsize\ttfamily,
  numbers=left,
  numberstyle=\scriptsize,
  numbersep=5pt,
  breaklines=true,
  breakatwhitespace=true,
  xleftmargin=1.5em
}
\begin{lstlisting}[%
  language=C,
  caption={OpenCL kernel and host code example},
  label={lst:opencl_example}%
]
__kernel void foo() {
    int tid = get_global_id(0);
    ...
    barrier(...);}

__host int main() {
    ...
    cl_mem A_clmem = clCreateBuffer(context, ...);
    clEnqueueNDRangeKernel(command_queue, foo, ...);
    ...
    return 0;}
\end{lstlisting}

\Cref{fig:design:framework} presents the overall structure of the \name{} framework. The framework has two compilation processes: a host compilation process and a kernel compilation 
process.
%: a host process and a kernel process 
As shown in \cref{lst:opencl_example}, a GPU program comprises host code and kernel code. 
During compilation, the front-end rewrites host-side API calls used for communication and buffer management (e.g., \texttt{clEnqueueNDRangeKernel()}, \texttt{clCreateBuffer()}) into runtime operations via the device runtime library. In addition, the compiler analyzes kernel code semantics that carry runtime information or program control flow (e.g., \texttt{get\_global\_id()}, \texttt{barrier()}) and generates a GPU binary. 

\subsection{Front-end Compiler}

The front-end compiler compiles both host and kernel code. During the host compilation process, it employs a runtime translation layer and links against the target device runtime library to interface with the device driver and runtime. In the kernel compilation process, the front-end compiler generates LLVM IR by analyzing language semantics, applying semantics-aware optimizations, and inserting thread-scheduling code. During this stage, math and atomic calls in the kernel code are resolved against the appropriate built-in libraries for the target hardware. 

\textbf{Semantics-aware code optimization}: 
\name{} extends memory structure handling and built-in libraries for Vortex.
This optimization leverages language semantics to apply targeted optimizations in three stages. First, the front-end compiler analyzes the memory usage of the program (shared, constant, and global) and either maps memory to the hardware. We extend memory conversion for the target open-GPU architecture (we will discuss memory semantics support in \cref{sec:extension2}).
Second, the optimization examines each function call and replaces kernel intrinsics (e.g., OpenCL’s \texttt{get\_local\_id()}) with explicit function parameters, since this information (e.g., \texttt{local\_id}, \texttt{local\_size}) is passed via arguments.
Third, the optimization finds special function calls (e.g., barrier, math, print, and atomic functions), then lowers each call appropriately in the built-in library. \name{} extend the built-in library to support Vortex.

\textbf{Thread-schedule code insertion}: 
Mapping kernel code onto open‑GPU hardware in a way that maximizes performance hinges on an efficient scheduler, yet exposing low‑level scheduling details to end users undermines portability and productivity. Unlike the GPU programming model (e.g., OpenCL’s work‑item/work‑group), GPU hardware adopts a thread/wavefront model in which a wavefront—a fixed‑size group of threads—executes in lock‑step on the GPU’s SIMD lanes. To bridge this semantic gap automatically, \name{} performs thread-schedule code insertion during kernel compilation: the front‑end compiler analyzes parallelizable code regions and, when necessary, wraps those regions in appropriate loops, thereby enabling high‑level work‑item semantics to execute as wavefronts and individual threads.

\subsection{Middle-end compiler}

The middle-end compiler applies target-independent optimizations and GPU-aware analyses, enabling reuse across GPU architectures. \name{} builds on LLVM~\cite{paper:llvm}, reusing existing passes (O2 optimization flag) and introducing new optimizations where needed. Especially, \name{} centralizes the divergence-management-related analyses and optimizations into the middle-end.

\begin{figure}[t]
    \centering
    \begin{subfigure}[b]{0.15\textwidth}
        \centering
        \includegraphics[width=\textwidth]{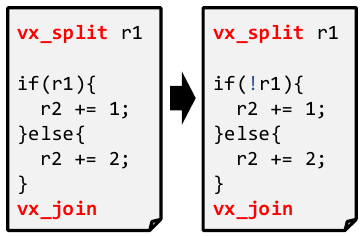}
        \caption{Reordering}
        \label{fig:design:ir:reordering}
    \end{subfigure}%
    \hfill
    \begin{subfigure}[b]{0.15\textwidth}
        \centering
        \includegraphics[width=\textwidth]{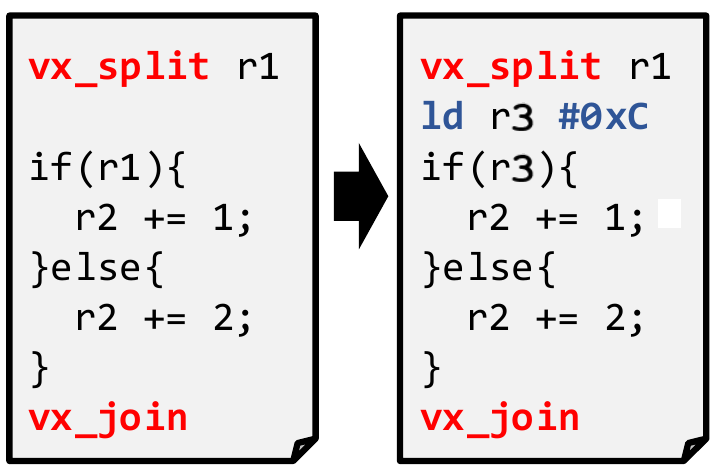}
        \caption{Spilling}
        \label{fig:design:ir:spilling}
    \end{subfigure}%
    \hfill
    \begin{subfigure}[b]{0.15\textwidth}
        \centering
        \includegraphics[width=\textwidth]{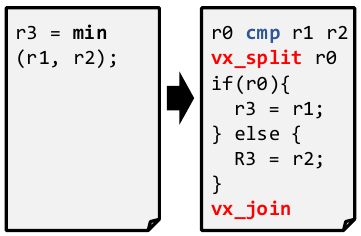}
        \caption{Simplification}
        \label{fig:design:ir:simplification}
    \end{subfigure}
    \caption{Challenges of Split/join IR insertion}
    \label{fig:divergence_challenges}
    \vspace*{-\textfloatsep}
\end{figure}

One of the most critical challenges in GPU compiler design is providing effective support for divergence control, analysis, and performance optimization. As discussed in \cref{sec:back:vortex_isa}, conditional branches and loops can lead to thread divergence under SIMT execution. Therefore, proper divergence-control instruction insertion and code transformation are essential. To ensure correctness and efficiency, the compiler incorporates thread divergence analysis along with optimizations that preserve program semantics and produce code that executes reliably on hardware, as suggested by prior works~\cite{cc:14:taming, arxiv:24:cf_management, caches:11:characterization, cgo:22:dram, isca:12:capri, hpca:13:dual-path, hpca:14:multipath, hpcs:16:cf_restructuring, ics:15:dacache, micro:14:exploringDSE, pact:11:divergence, pact:12:data_herding, popl:03:folklore, micro:07:ipdom, arxiv:24:cf_management}.

Managing divergence is nontrivial when instrumentation is inserted at the IR level,
because the CFG can still be modified by later back-end passes.
\Cref{fig:divergence_challenges} summarizes three correctness hazards that can occur.
First, branch reordering (\Cref{fig:design:ir:reordering}) performed after IR emission can invalidate previously inserted divergence sites:
swapping then/else blocks or inverting the condition without updating the
\textbf{vx\_split} predicate activates the incorrect lanes.
Second, spilling/reloading (\Cref{fig:design:ir:spilling}) can cause
\emph{predicate drift}: the back-end may reload the branch predicate into a new vreg (e.g., \texttt{r3})
while \textbf{vx\_split} still references the original vreg (e.g., \texttt{r1}).
Third, IR simplification (\Cref{fig:design:ir:simplification}) of a divergent \emph{select}
may later be expanded to a compare-and-branch in the back-end, thereby skipping the required
\{\textbf{vx\_split}, \textbf{vx\_join}\} instrumentation.

To make divergence management both portable and maintainable, \name{} performs nearly all divergence planning and insertion in the target-independent LLVM IR, and then applies a
\emph{lightweight} MIR safety net as the \emph{last} Machine IR pass after
register allocation. Concretely: (a) for late branch inversions, \name{} simply flip the \emph{negate} flag on the corresponding \textbf{vx\_split} so that lane semantics align. (b) To eliminate
predicate drift, \name{} unify \textbf{vx\_split}'s operand with the machine branch predicate, moving them back-to-back. (c) For
divergent \texttt{select}, \name{} reify it as a diamond CFG \emph{in the IR} and
inserts \{\textbf{vx\_split}, \textbf{vx\_join}\}.
Together, these measures keep the MIR pass minimal while ensuring predicate
consistency, split/join pairing, token validity, and mask restoration under
last-phase CFG changes—preserving portability across targets and maintainability across compiler updates.

\ignore{
\begin{figure}[t]
    \centering
    \includegraphics[width=0.48\textwidth]{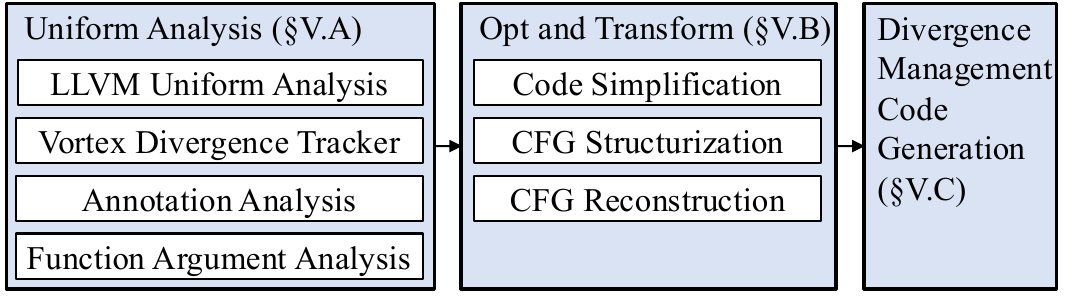}
    \caption{Overview of Divergence Management in \name{}}
    \label{fig:divergence_management}
    \vspace*{-\textfloatsep}
\end{figure}
}

% divergence tracker + uniform annotation + uniform analysis 0 
\subsubsection{Uniformity Analysis}

Divergence management mechanisms begin with uniformity analysis. Uniformity analysis determines whether a program variable maintains the same value across all threads in a warp/wavefront (uniform) or whether its value may vary across threads (divergent). This distinction is critical because the behavior of control-flow constructs, such as branch or select, depends directly on the uniformity of their controlling values. If the condition is uniform, all threads within a warp follow the same execution path, enabling efficient execution. In contrast, if the condition is divergent, threads may take different paths, forcing the hardware or compiler to manage divergent execution explicitly by indicating divergence and reconvergence points.

\textbf{Extending LLVM Uniform Analysis}: Uniformity analysis of \name{} is build on top of uniform analysis of LLVM by extending abstract interface.
Uniform analysis of LLVM first identifies seed information by marking values that are always-uniform (such as constant values) and values that are sources of divergence (such as thread identifiers). From these initial values, the analysis propagates uniformity information along the program by using def–use chains and through control-dependence relationships. For example, a divergent branch condition marks all instructions that are control-dependent on it as potentially divergent. To support target-specific divergence management, LLVM uses the Target Transformation Info (TTI) interface, which stores transformation information for each target. This interface includes functions such as \texttt{isAlwaysUniform} and \texttt{isSourceOfDiver-} \texttt{gence}, allowing the compiler to generate initial values in a target-specific manner.

RISC-V was originally designed for CPUs, and therefore the llvm-riscv back-end does not consider branch divergence. To address this limitation, \name{} extends the RISC-V TTI interface with divergence management by adding the functions \texttt{isAlwaysUniform} and \texttt{isSourceOfDivergence}, thereby enabling LLVM uniform analysis to serve as the default basis for divergence analysis.

\textbf{\name{} Divergence Tracker}: To implement these two functions, \name{} proposes a divergence tracker. The divergence tracker identifies \textit{sources of divergence} under the following conditions. First, it conservatively assumes that all function arguments are potentially divergent except when they are marked as uniform. Second, atomic operations are considered divergent because their results may differ across threads. When multiple threads perform an atomic operation on the same memory location, each thread observes a different result depending on the order of execution. 
Third, it conservatively assumes that function return values are divergent if they are not marked as uniform in \texttt{isAlwaysUniform}; this also includes intrinsics such as \texttt{threadlocal\_address} used for accessing TLS variables.

\textit{Always-uniform variables} are determined based on RISC-V Control and Status Register(CSR). First, all machine-level CSRs are assumed to be always uniform since their value is uniform across all threads (e.g. \texttt{num\_threads}, \texttt{num\_warps}), and custom user-level CSRs such as \texttt{core\_id}, \texttt{warp\_id} that are always uniform. Second, function return values are uniform if they are marked as such.

\textbf{Annotation Analysis}: To complement the base uniformity analysis, \name{} incorporates annotation-based reasoning. This analysis operates at two levels. First, intrinsic-based analysis examines global variables, constant data variables, and stack variables, where store and load operations to stack variables are conservatively marked as uniform. Second, uniform information can also be derived from user-provided annotations. In this step, the analysis processes both metadata-based annotations (e.g., “vortex.uniform”) and intrinsic-based annotations to ensure consistency.

\textbf{Function Argument Analysis}: \name{} extends uniformity analysis to function
arguments, which are typically treated conservatively as divergent. \name{} build the
call graph and run our function-level analysis in \emph{reverse post-order}
(RPO) to determine the uniformity of each function’s arguments and return value (see \cref{alg:arg-analysis}).
\name{} honor explicit annotations (uniform/divergent) and analyze pointer arguments
for potential non-uniform accesses. After convergence, if a function has
internal linkage and an argument is proven uniform, that parameter is marked
\emph{uniform}.

Together, these extensions improve the precision of uniformity analysis and strengthen the foundation for reliable divergence management.

\begin{algorithm}[t]
\caption{Function Argument Analysis}
\label{alg:arg-analysis} 
\footnotesize
\DontPrintSemicolon
\SetKwFunction{BuildCG}{BuildCallGraph}
\SetKwFunction{SCCOrder}{SCCOrder}
\SetKwFunction{RPO}{ReversePostOrder}
\SetKwFunction{Analyze}{AnalyzeFunc}
\SetKwFunction{AnalyzeArgument}{AnalyzeArgument}
\SetKwFunction{AnalyzePtrOut}{AnalyzePtrOut}
\SetKwFunction{AnalyzeRetVal}{AnalyzeRetVal}
\SetKwProg{Fn}{Procedure}{:}{end}
\KwIn{Module $M$}
\KwOut{$\mathsf{UArg}[f,i]$, $\mathsf{UPtrOut}[f,i]$, $\mathsf{URet}[f]$}
$CG \gets$ \BuildCG{$M$}\;
$\mathcal{F} \gets$ \RPO{\SCCOrder{$CG$}}\;
\ForEach{$f \in \mathcal{F}$}{
  initialize $\mathsf{UArg}[f,*], \mathsf{UPtrOut}[f,*], \mathsf{URet}[f] \gets \textsc{Divergent}$
}
\Repeat(){no change}{
  \ForEach{$f \in \mathcal{F}$}{
    \Analyze{$f$}
  }
}
\Fn{\Analyze{$f$}}{
  $\mathsf{UArg}   \;\leftarrow\; \mathsf{UArg}   \;\cup\; \AnalyzeArgument{f}$\;
  $\mathsf{UPtrOut}\;\leftarrow\; \mathsf{UPtrOut}\;\cup\; \AnalyzePtrOut{f}$\;
  $\mathsf{URet}   \;\leftarrow\; \mathsf{URet}   \;\cup\; \AnalyzeRetVal{f}$\;
}
\Return{$(\mathsf{UArg}, \mathsf{UPtrOut}, \mathsf{URet})$}\;
\end{algorithm}

\begin{algorithm}[t]
\caption{CFG Transformation}
\label{alg:divergence-aware-cfg}
\footnotesize
\DontPrintSemicolon
\KwIn{Original CFG $G$}
\KwOut{Transformed CFG $G'$}

\SetKwFunction{Proc}{Transform\_Func}
\SetKwFunction{IsUniform}{IS\_UNIFORM}
\SetKwFunction{IsConditional}{IS\_CONDITIONAL}
\SetKwFunction{IsLoopBranch}{IS\_LOOP\_BRANCH}
\SetKwFunction{IPDom}{IPDOM}
\SetKwFunction{Reach}{IS\_REACHABLE}
\SetKwFunction{TLoop}{TRANSFORM\_LOOP}
\SetKwFunction{TBr}{TRANSFORM\_BRANCH}
\SetKwProg{Fn}{Procedure}{:}{}

\Fn{\Proc{$G$}}{
  $\mathcal{D}_{\text{branch}} \leftarrow \emptyset$ \;
  $\mathcal{D}_{\text{loop}} \leftarrow \emptyset$ \;

  \ForEach{branch $b$ in $G$}{
    \If{\IsUniform{$b$} \textbf{or} $\neg$\IsConditional{$b$}}{
      \textbf{continue} \;
    }
    $\mathit{ip} \leftarrow FindIPDom(b)$ \;

    \If{\IsLoopBranch{$b$}}{
      \If{$\mathit{ip}$ is inside the loop of $b$}{
        $\mathcal{D}_{\text{branch}} \leftarrow \mathcal{D}_{\text{branch}} \cup \{(b,\mathit{ip})\}$ \;
      }\Else{
        $\mathcal{D}_{\text{loop}} \leftarrow \mathcal{D}_{\text{loop}} \cup \{(b,\mathit{ip})\}$ \;
      }
    }\Else{
        \If{IS\_REACHABLE(b, ip)}{
        $\mathcal{D}_{\text{branch}} \leftarrow \mathcal{D}_{\text{branch}} \cup \{(b,\mathit{ip})\}$ \;
        }
    }
  }
 $G \leftarrow$ \TLoop{$G$, $\mathcal{D}_{\text{loop}}$}\;

 $G \leftarrow$ \TBr{$G$, $\mathcal{D}_{\text{branch}}$}\;
  \textbf{return} $G$\;
}
\end{algorithm}

\begin{figure}[t]
    \centering
    \includegraphics[width=0.48\textwidth]{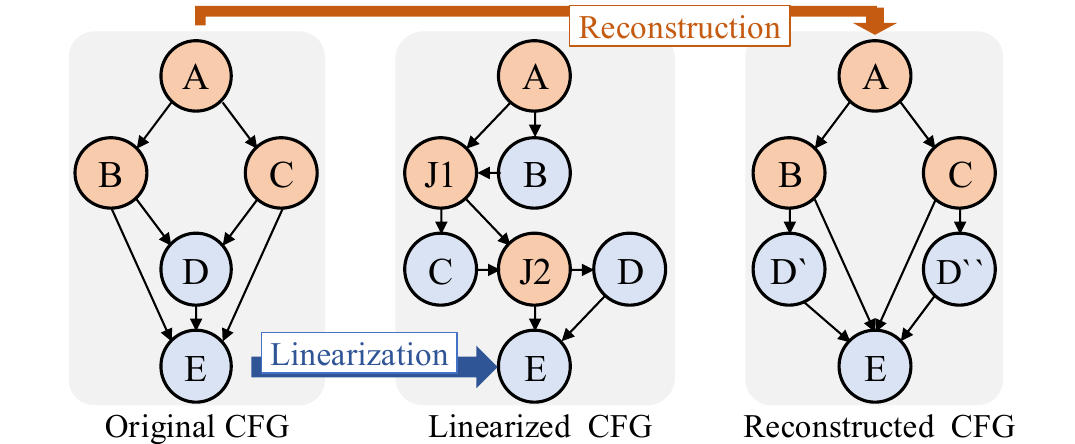}
    \caption{CFG Structurization and CFG Reconstruction}
    \label{fig:opt:cfg}
    \vspace*{-\textfloatsep}
\end{figure}

% =================================================== 
% contents in Vortex Divergence0 + zicond
\subsubsection{Optimization and Transformation}

%\textbf{Code Simplification}:
%\textbf{Code and CFG Simplification}: \name{} simplifies control flow and canonicalizes functions into a single-exit form while removing unreachable blocks. First, \name{} simplifies the code by normalizing away \texttt{select}, \texttt{min}, \texttt{max}, and integer conditional operations, rewriting them into equivalent branch\-based control flow while preserving semantics. When the target machine provides native support for these operators, simplification is skipped and the IR is lowered directly to the corresponding instructions. In particular, \name{} implements an option to enable replacing integer conditional operations (ZiCond) with \texttt{CMOV}. Second, \name{} runs LLVM code-simplification passes and applies additional CFG simplification to ensure a single exit block by folding functions with multiple \texttt{return} instructions into one unified exit block, and removes unreachable blocks.

\textbf{Code and CFG Simplification}: \name{} simplifies control flow and canonicalizes functions into a single-exit form while removing unreachable blocks. It first normalizes \texttt{select}, \texttt{min}, \texttt{max}, and integer conditional operations by rewriting them into equivalent branch-based control flow, unless the target machine provides native support. In that case, \name{} skips rewriting and directly lowers these operations, with an option to replace integer conditionals (ZiCond) with \texttt{CMOV}. \name{} then applies LLVM code-simplification passes and additional CFG cleanup to merge functions with multiple \texttt{return} instructions into one exit block and eliminate unreachable blocks.

\textbf{Control-Flow Structurization}:
For effective GPU divergence management, the control-flow graph (CFG) should be structured (reducible). A reducible flow graph is one whose edges can be uniquely partitioned into forward edges and back edges with the following properties: (1) removing all back edges yields an acyclic flow graph, and adding any single back edge reintroduces a cycle; and (2) for each back edge $n \to m$, node $m$ dominates $n$~\cite{paper:cfg:reducible}. An irreducible flow graph is simply one that is not reducible~\cite{paper:cfg:irreducible}. In SIMT execution, explicit reconvergence points are required so that the compiler can determine where threads rejoin after a branch and correctly apply thread-mask–based execution (logical predicates/thread masks). A structured CFG provides stable join points such as the immediate post-dominator (IPDOM). These join points determine where to place split/join constructs and where to save and restore thread masks. \name{} uses 
an LLVM pass \texttt{llvm::createStructurizeCFGPass()}.

\textbf{CFG Reconstruction}: However, when unstructured regions are deeply nested, predicate computation for linearization can become expensive. To mitigate this overhead, \name{} introduce a new optimization, CFG reconstruction, which selectively duplicates nodes to simplify predicates. As shown in \cref{fig:opt:cfg}, \name{} duplicate node \texttt{D} to \texttt{D'} and \texttt{D''}. This transformation is effective for graphs with complex control dependencies. For example, the OpenCL \texttt{cfd} benchmark’s control-dependence graph(CDG) exhibits substantial depth that can inflate linearization cost. 
An interesting observation is that if the dependency is uniform, each warp requires only a single pass, eliminating the need for duplication
Accordingly, when \name{} encounter an unstructured block such as node \texttt{D''} and the node is a divergent CDG leaf node, \name{} duplicate that node to reduce predicate complexity. %, as shown in \cref{alg:cfg-reconstruction}.

% contents in Vortex Divergence 1 + 2
\subsubsection{Divergence Management Function Insertion}

\Cref{alg:divergence-aware-cfg} shows the divergence management function insertion. First, the algorithm iterates over all branch instructions in the original CFG and first skipping those that are either uniform or non-conditional. Second, the algorithm finds the immediate post\-dominator (\textit{ip}) of branch \textit{b}. 
If the branch belongs to a loop, the algorithm distinguishes between divergence blocks and divergent loops depending on whether \textit{ip} lies inside the loop or branches for the loop structure. Otherwise, non-loop branches are checked for reachability to their post-dominator before being classified as divergent. 
After classification, \name{} invokes \texttt{TRANSFORM\_LOOP} and \texttt{TRANSFORM\_BRANCH} to process divergent loops and bran- ches, respectively, applying the appropriate divergence-hand- ling transformations.
This separation between detection and transformation ensures a systematic and structured approach to divergence management in the CFG.

\textbf{Loop predicate Insertion} (\texttt{TRANS\_LOOP}): For loop branches, divergence is managed through the use of predicate ISA and thread masks. A thread mask is inserted at the loop header and stored in the preheader to record the active threads entering the loop. At the loop exit, this mask is restored to maintain consistent control flow. This requires transformation to a single-exit-block loop, ensuring well-defined reconvergence. Additionally, a predicate is introduced to mask out threads that terminate early within the loop. %Finally, the branch condition is rewritten to enable divergence-aware optimization.

\textbf{Split and Join Insertion} (\texttt{TRANSFORM\_BRANCH}): For non\-loop branches, divergence is handled using a split and join ISA.
At the branch point, a split is inserted to separate divergent thread paths.
The branch condition is then modified to reflect the optimization requirements. Before the immediate post\-dominator, a join operation is inserted to merge the divergent paths back into a unified control flow, thereby ensuring correctness while limiting divergence overhead.

%This result suggests that we can pursue new optimizations, such as using some NOP instructions, to control memory operation density.

%9	reconstruction

%\subsection{Instruction Insertion}
% - split and join / ipdom control ? 

% - CMOV instruction  

%\SJ{Maybe it would be good to write about old methods(vortex 1.0) -> and use it as the baseline of Eval, and show our method's rationality}

%Control-flow Divergence Goals
%Preserving existing RISC-V branch ISA
%Branch optimizations are critical to GPU performance 
%Branch implement in LLVM is tightly coupled
%Affects instruction selection, peephole optimization passes
%Backend optimizations  
%One change ripples through the whole chain

%Split/join IR insertion challenges
%Branch reordering / inversion
%Register spilling
%Branch simplification
%Branch folding
%Jump threading
%Critical‐edge splitting/unsplitting

%\textbf{Solution: Raise thread divergence analysis in the middle-end compiler.}
%We move IR to the standard IR ( middle-end IR level) 
%    - (improve flexibility) can cover future IR optimization 
%    - (improve portability) reuse divergence result in different hw or ISA

\subsection{Back-end Compiler}

The back‑end compiler handles target‑specific optimizations and final code generation. \name{} extended the LLVM RISC‑V target compiler to generate target binaries for the Vortex GPU. 
During binary generation, the compiler takes hardware-specific constraints into account, enabling safe and automatic use without exposing these constraints to higher layers.

\textbf{ISA table extension}: \name{} extends the LLVM RISCV ISA description table used in code generation, adding support for Vortex instructions and intrinsics.

\textbf{Vortex target code generation}: 
In the back-end compiler, the device kernel lowering stage takes the fully optimized IR and lowers it to Vortex instructions, linking the Vortex Kernel library to produce a complete kernel executable. A final machine‑code optimization pass then eliminates redundant register‑copy instructions and corrects divergence split/join pairs to ensure a streamlined, divergence‑safe binary.

%\textbf{Divergence management}: 
%Control‑flow divergence handling in \name{} is achieved through a two‑step process. First, the analysis pass examines each kernel’s control‑flow graph to identify and mark every divergent instruction and region. Second, the management pass inserts explicit Split and Join instructions around those points, as illustrated in \cref{fig:design:ex}. For simple if/else branches, split instructions appear at the branch and join instructions after both paths; for loops—where threads may enter and exit the body at different iterations—split/join pairs are injected at loop entry and exit, and the predicate instruction inside the loop body dynamically disables or re‑enables threads according to the loop condition. This combination of divergence detection, structured split/join insertion, and predicate‑based mask updates allows \name{} to systematically control both branch and loop divergence, preserving correctness and delivering high‑performance SIMD execution.

%\subsection{compile flags}

%To support GPU-aware optimization and code generation, \name{} 

%% file: paper/3.2.Eval.tex
\section{Evaluation}

In the evaluation, we conducted our experiments using SimX \cite{vortex_first_2020}.
SimX provides deterministic, cycle-accurate execution (within 6\% of RTL), ensuring that repeated runs yield identical results and that performance differences arise solely from compiler transformations rather than simulator noise.
All experiments were performed under a 4-core, 16-warp, 32-thread configuration with L2 cache enabled, reflecting the standard architectural parameters used in prior Vortex evaluations.
Correctness is validated by comparing all benchmark outputs against reference CPU implementations

\subsection{Benchmark Coverage}
\label{sec:eval:coverage}

To evaluate the coverage of \name{}, we tested 32 OpenCL benchmarks and 17 CUDA benchmarks, verifying correctness for all supported workloads. Specifically, \name{} supports 12 benchmarks from the NVIDIA OpenCL SDK~\cite{benchmark:nvidia_openCL} (convolution, DotProduct, gaussian, oclprintf, psort, reduce0, saxpy, sfilter, transpose, vecadd, VectorHypot, blackscholes, psum), 7 benchmarks from Parboil~\cite{benchmark:parboil} (cutcp, lbm, mri-q, sad, sgemm, spmv, stencil), and 13 benchmarks from Rodinia~\cite{benchmark:rodinia} (nearn, bfs, BlackScholes, kmeans, gaussian, pathfinder, backprop, hotspot3d, b+tree, lavaMD, myocyte, srad, cfd). For CUDA, \name{} additionally runs 12 Rodinia benchmarks (Gauss, bfs, nn, nw, srad\_v2, hotspot, StreamCluster, Myocyte, Pathfinder, LUD, Btree, Backprop) and 5 benchmarks from HeCBenchmark~\cite{kernel_fission_benchmark_suite} (vote-cuda (Vote), shuffle-cuda, bscan-cuda (Bscan), atomic-aggregate (AtomicAgc), gc-cuda (GC)). These benchmarks collectively exercise key SIMT behaviors, including divergence, shared-memory usage, and compute-intensive kernels, providing broad coverage across GPU execution patterns.

\subsection{Effect of Divergence Management}

\begin{figure}[t]
\centering
    \includegraphics[width=0.48\textwidth]{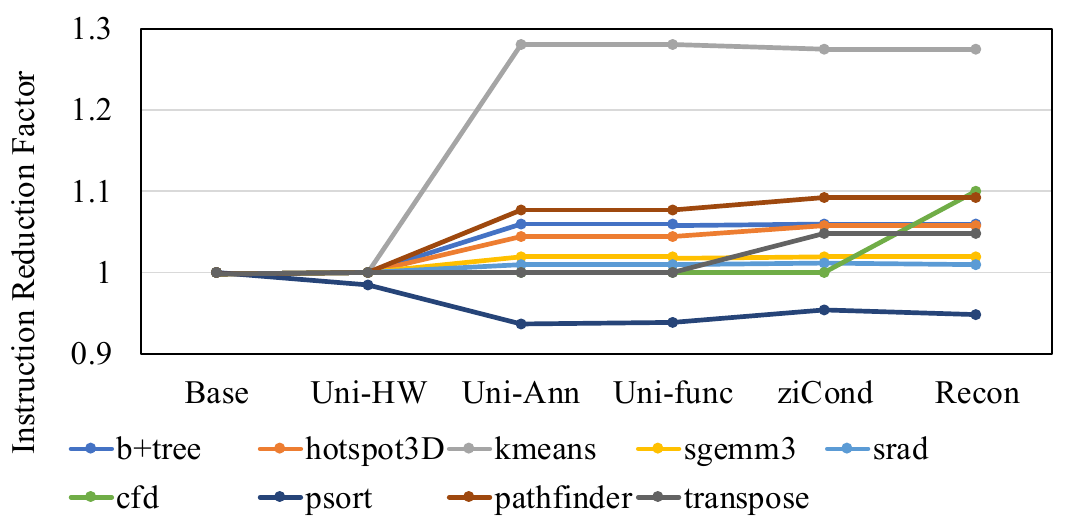}
    \caption{Instruction Reduction by Divergence Optimization}
    \label{fig:ext1:graph:inst}
    %\vspace{-\textfloatsep}
    \vspace{-5mm}
\end{figure}

\begin{figure}[t]
\centering
    \includegraphics[width=0.48\textwidth]{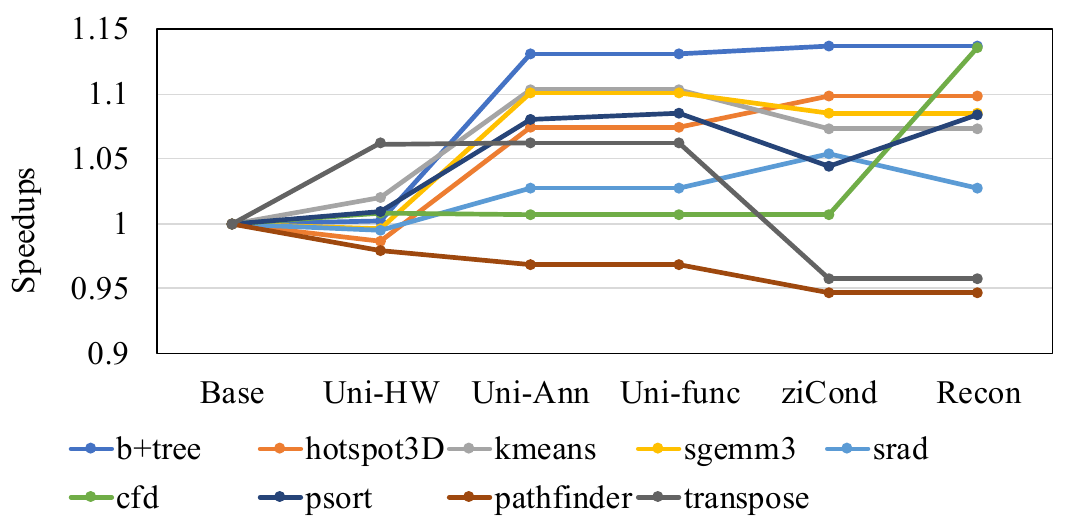}
    \caption{Speedups by Divergence Optimization}
    \label{fig:ext1:graph:speedups}
    %\vspace{-\textfloatsep}
        \vspace{-5mm}
\end{figure}

This evaluation measures the performance impact of using uniformity analysis and code optimizations. 
First, the baseline includes all the analyses and optimizations required for the code to work correctly on Vortex(source Of Divergence analysis, divergence tracker, code simplification, control-flow structures, CFG linearization, divergence management function insertion). Second, our uniformity analysis includes always-uniform value analysis based on hardware structure in divergence tracker (Uni-HW), annotation analysis (Uni-Ann), and function argument analysis (Uni-Func). Lastly, we apply optimizations such as lowering the ternary operator to the CMOV instruction of Vortex (ZiCond) and CFG reconstruction (Recon). \Cref{fig:ext1:graph:inst} and \cref{fig:ext1:graph:speedups} report only the benchmarks with noticeable changes; the remaining benchmarks listed in \cref{sec:eval:coverage} show negligible differences.

\Cref{{fig:ext1:graph:inst}} shows the instruction reduction factor (higher is better). By adding analyses and optimizations, we observe that the number of instructions is generally reduced. Especially, Uni-Ann shows a clear impact on instruction reduction. Since not all benchmarks contain ternary operators and irreducible CFGs, some benchmarks show performance improvement with the ZiCond optimization (grad, pathfinder, and transpose) and the CFG reconstruction (CFD). However, not all benchmarks benefit from divergence analysis, as it can also increase the instruction count, such as in psort.

\Cref{{fig:ext1:graph:speedups}} shows the speedup from the analyses and optimizations. As shown, divergence analysis also brings speedups. In particular, the annotation pass is important in terms of speedups. CFD also gains performance with the reconstruction pass. One interesting observation is that ZiCond optimization can reduce the instruction count for pathfinder and transpose but increase the memory request density,  thereby causing slowdowns. 

The combination of tested optimizations incurs only a 0.18\% compile-time geomean slowdown and introduces negligible binary-size overhead. Given that our approach operates in O(n) time, we expect it to scale even for large codebases.

%% file: paper/6.case_study_arch.tex
\subsection{Case Study 1: Extending ISA support}
\label{sec:extension3}

In this section, we discuss how \name{}’s support can be extended to other GPU ISAs. We begin by examining how \name{} integrates ISA extensions for Vortex, and then show how the same approach generalizes to other open GPU projects.

To improve hardware performance, introducing new microarchitectural features is common. In such cases, the compiler can connect front-end languages with the new ISA in two approaches: (i) the ISA can be added to the ISA table of back-end compiler, so that LLVM IR can be translated into ISA instructions or (ii) when the front-end compiler provides a semantically equivalent gpu-specific function (e.g., CUDA warp shuffle), the front-end compiler can replace the function call with a built-in function implemented using inline assembly of the ISA. This approach is easy to implement, making it a suitable choice when the function is not optimized or has been altered by compiler optimization. 

The CMOV instruction is one example of the first approach. While it is true that this instruction is isomorphic to the ternary operator and thus can be mapped one-to-one, it can also be applied in cases where code is written with an if-else construct but still benefits from performance improvement when expressed as a CMOV instruction. Consequently, the compiler may introduce CMOV during optimization. Therefore, CMOV was added to \name{} via the first approach, with the \textbf{vx\_move} being incorporated into the ISA table.

The second approach applies to cases that support warp-level features such as warp voting and shuffling. These functions already appear in CUDA as warp-level functions that closely match the hardware ISA. Since their execution is hardware-accelerated~\cite{dateworkshop:25:pu}, it is sufficient to translate them directly into hardware instructions using the built-in library.

In more detail, we modify the CuPBoP to identify and replace warp-level function calls in the built-in library using Vortex instructions. Warp vote and warp shuffle in the CUDA are compiled into NVVM-specific intrinsics. The compiler scans over the code for these intrinsic calls and substitutes them with Vortex intrinsics \textbf{vx\_shfl} and \textbf{vx\_vote}, respectively. 
A key challenge is that Vortex intrinsics definitions are not available during the IR transformation. As vortex intrinsics are defined in the header file as inline functions, the definitions cannot be compiled and linked to the transformed program. Another challenge is the interface difference between the CUDA and Vortex intrinsics. To overcome this, we extended CuPBoP runtime to implement warp functions with Vortex intrinsics. The CUDA intrinsics are replaced with CuPBoP runtime functions instead. 

\begin{figure}[t]
    \centering
    \includegraphics[width=0.48\textwidth]{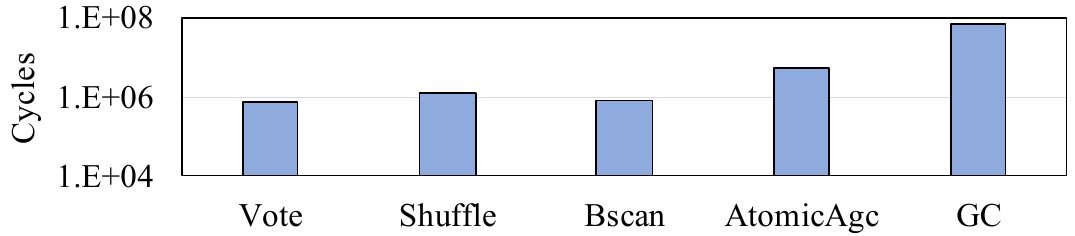}
    \caption{Performance of ISA  extension on Vortex GPU by supporting vote, move, shuffle, atomic ISAs}
    \label{fig:eval3}
    %\vspace*{-\textfloatsep}
    \vspace{-2mm}
\end{figure}

%\cref{fig:eval3} shows the performance of the CUDA benchmarks from HeCBenchmark~\cite{kernel_fission_benchmark_suite}. 
%By extending the ISA and CuPBoP, we can now successfully support new instructions and benchmarks such as vote-cuda(Vote), suffle-cuda, bscan-cuda(Bscan), atomic-aggregate(AtomicAgc), and gc-cuda(GC).

By extending CuPBoP, \name{} can support more benchmarks as show in \cref{fig:eval3}.

%% file: paper/5.case_study_frontend.tex
\begin{figure}[t]
    \centering
    \includegraphics[width=0.48\textwidth]{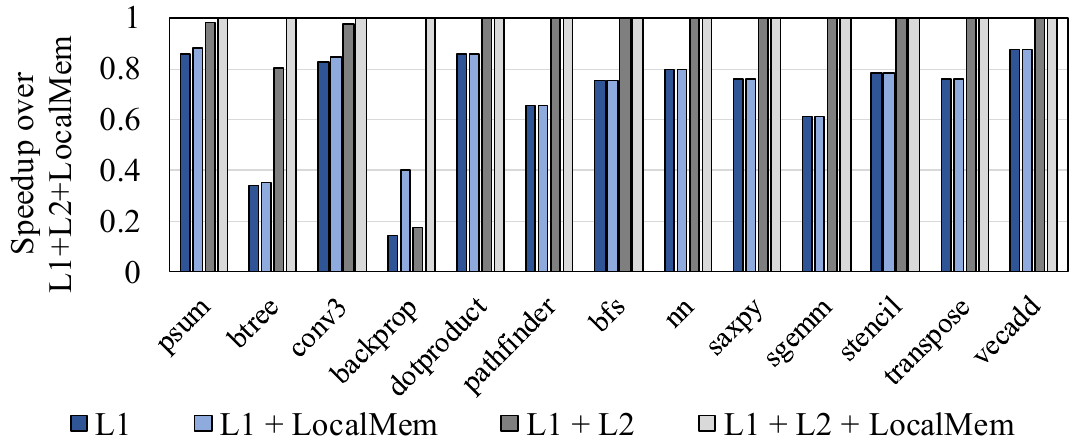}
    \caption{Performance effect with different cache configurations as well as different ways of supporting shared memory}
    \label{fig:eval2}
    %\vspace*{-\textfloatsep}
\end{figure}

\subsection{Case Study 2: Extending Host Memory APIs}
\label{sec:extension2}

CuPBoP~\cite{fccm:25:softcuda} maps CUDA constant and shared memory to Vortex global memory to preserve correctness, but it does not implement CUDA’s memory-related host APIs such as \texttt{cudaMemcpyToSymbol}. As a result, CUDA applications that rely on these APIs cannot properly initialize device-side variables on Vortex.

\name{} extends this capability through a modular host–runtime design that allows new API semantics to be incorporated without modifying the core compiler pipeline. To support \texttt{cudaMemcpyToSymbol}, which is widely used to initialize \textit{constant} variables on NVIDIA GPUs, \name{} reinterprets constant memory semantics under the Vortex memory model. Constant variables are lowered to global memory, and their initialization is emulated in software. When the host invokes the API, data is buffered on the device and materialized just before kernel launch,
after global addresses are resolved. This approach preserves CUDA semantics while requiring no changes to existing applications.

\name{} generalizes this extensibility to CUDA shared memory as well. While CuPBoP lowers all shared memory to global memory, \name{} additionally supports mapping shared memory onto Vortex’s per-core local memory. As shown in \cref{fig:eval2}, the choice of mapping can influence performance depending on the kernel's computation-to-memory ratio and shared-memory usage
patterns.

These extensions demonstrate that \name{}’s host–runtime layer is not fixed to a single memory semantics model, but instead provides a flexible and extensible interface that can accommodate API behaviors and memory mappings required by different open GPU architectures.

%% file: paper/7.related_works.tex
%\section{Related Work}

%\subsection{Industry Directions of handling divergence handling}

%\textbf{AMD}

%\textbf{NVIDIA}

%\textbf{Others}

%\subsection{Different compiler approaches of divergence handling}

%% file: paper/8.discussion.tex
\section{Discussion}

\subsection{Extending support for open GPU variants}
%\name{} enables multiple integration paths depending on the needs of a given GPU project. Instead of requiring wholesale adoption, researchers can selectively integrate \name{} components at different levels of the stack. We identify the following extension paths:

Our two case studies illustrate the primary axes along which GPU extensibility typically evolves: ISA-level changes and runtime or memory-semantics changes.
Building on this observation, \name{} supports multiple integration paths depending on the needs of a given GPU project.
Rather than requiring wholesale adoption, researchers can selectively integrate \name{} components at different layers of the stack. We identify the following extension paths:

\textbf{Full Stack Integration:}  
  Projects adopt \name{}’s runtime libraries, front-end support (OpenCL/CUDA), middle-end passes, 
  and back-end ISA extensions. This path is best suited for Vortex-derived designs (e.g., Virgo~\cite{asplos:25:vringo}, SparseWeaver~\cite{vortex:sparseweaver}), where the SIMT execution model and divergence instructions closely resemble 
  Vortex. It provides maximum portability and allows existing CUDA/OpenCL applications to run with 
  minimal modification.

\textbf{Compiler Framework Extension:}  
  \name{}’s compiler stack is reused while plugging into an existing or custom runtime. 
  It is suitable for projects that already provide a runtime API but lack portable compiler passes. 
  For example, Ventus can reuse \name{}’s uniformity analysis and divergence-aware transformations 
  while keeping its own runtime system.

\textbf{Middle-End Analysis:} \name{}’s uniformity analysis, CFG structurization, and divergence handling passes are 
  integrated into an existing LLVM pipeline. This path fits vector-based RISC-V GPUs 
  (e.g., Ventus~\cite{iccd:24:ventus}, Simty~\cite{carrv:simty}) that do not share the Vortex ISA but can reuse analysis infrastructure.

\textbf{Selective Feature Extension.}  
  \name{}’s mechanisms for exposing warp-level features (e.g., shuffle, vote, atomics) can be adopted 
  incrementally, without requiring full compiler integration. This approach is useful for projects 
  seeking to expose specific features to programmers while retaining their existing toolchains.

\textbf{Front-End Language Reuse.}  
  Projects focused primarily on enabling new programming models (e.g., CUDA $\rightarrow$ LLVM IR 
  translation via CuPBoP, or OpenCL via PoCL) can adopt \name{}’s front-end infrastructure while 
  maintaining their own middle-end and backend compilers. This enables language portability without 
  deep compiler integration.

\subsection{Maturity of \name}
VOLT is currently implemented as a research prototype and is being used internally.
In addition to support benchmark in the \cref{sec:eval:coverage}, \name{} can also generate binary for several published GPU variants, demonstrating its capability in graph processing applications (PR, BFS, SSSP, CC) and in code generation for GEMM and FlashAttention3 in SparseWeaver~\cite{vortex:sparseweaver} and Virgo~\cite{asplos:25:vringo}, respectively. 
As future work, \name{} can extend to support ML and LLM, building on ongoing research toward adding eventual Tensor Core.

%% file: paper/9.conclusion.tex
\section{Conclusion}

%We present \name{}, a compiler framework for open-source GPUs. \name{} supports OpenCL and CUDA as front-ends and targets RISC-V–based GPU implementations, with the Vortex project as our primary hardware target. In this paper, we showcase two case study to illustrate how host-API support and ISA extension can be supported.

This paper provides the first tool-oriented description of \name{}, presenting its design principles, overall structure, and centralized SIMT-aware compiler transformations. \name{} supports OpenCL and CUDA front-ends, centralizes divergence-management analyses in the middle-end, and demonstrates extensibility through host–runtime API and ISA extension case studies.
%-------- 

%\begin{tabular}{lll}
%\toprule
%\textbf{} & \textbf{Item} & \textbf{Status} \\
%\midrule
%\multirow{6}{*}{Writing} 
%    & abst & draft \\
%    & intro & draft\\
%    & back & draft \\
%    & design & \\
%    & opt & \\
%    & eval & \\
%\midrule
%\multirow{8}{*}{Evaluation} 
%    & extending benchmark & 31 \\
%    & 0. baseline & Done \\
%    & 1. 0 + vortex::uniformannotation & Done \\
%    & 2. 1 + zicond & Done \\
%    & 3. 2+ cross\_function\_opt & Done \\
%    & 4. 3 + reconstruction pass & \\
%    & 5. 4 + user annotation & \\
%\bottomrule
%\end{tabular}

%Evaluation Result :%https://shorturl.at/HD7CW

%% file: paper.bib
@misc{vortex_first_2020,
  title = {Vortex:OpenCL Compatible RISC-V GPGPU},
  author = {Elsabbagh, Fares and Tine, Blaise and Roshan, Priyadarshini and Lyons, Ethan and Kim, Euna and Shim, Da Eun and Zhu, Lingjun and Lim, Sung Kyu and Kim, Hyesoon},
  year = {2020},
  publisher = {{arXiv}},
}

@inproceedings{vortex:skybox,
author = {Tine, Blaise and Saxena, Varun and Srivatsan, Santosh and Simpson, Joshua R. and Alzammar, Fadi and Cooper, Liam and Kim, Hyesoon},
title = {Skybox: Open-Source Graphic Rendering on Programmable RISC-V GPUs},
year = {2023},
booktitle = {Proceedings of the 28th ACM International Conference on Architectural Support for Programming Languages and Operating Systems},

}

@inproceedings{vortex,
author = {Tine, Blaise and Yalamarthy, Krishna Praveen and Elsabbagh, Fares and Kim, Hyesoon},
title = {Vortex: Extending the RISC-V ISA for GPGPU and 3D-Graphics},
year = {2021},
booktitle = {MICRO-54: 54th Annual IEEE/ACM International Symposium on Microarchitecture},
}

@INPROCEEDINGS{vortex:ipdpsw,
  author={Ahn, Chihyo and Jeong, Shinnung and Cooper, Liam Paul and Parnenzini, Nicholas and Kim, Hyesoon},
  booktitle={2024 IEEE International Parallel and Distributed Processing Symposium Workshops}, 
  title={Comparative Analysis of Executing GPU Applications on FPGA: HLS vs. Soft GPU Approaches}, 
  year={2024},
}

@INPROCEEDINGS{vortex:sparseweaver,
  author={Jeong, Shinnung and Coopert, Liam Paul and Lee, Ju Min and Choi, Heelim and Parnenzini, Nicholas and Ahn, Chihyo and Lee, Yongwoo and Kim, Hanjun and Kim, Hyesoon},
  booktitle={2025 IEEE International Symposium on High Performance Computer Architecture (HPCA)}, 
  title={SparseWeaver: Converting Sparse Operations as Dense Operations on GPUs for Graph Workloads}, 
  year={2025},
  volume={},
  number={},
  pages={1437-1451},
  doi={10.1109/HPCA61900.2025.00108}
}

@inproceedings{cc:14:taming,
author="Anantpur, Jayvant
and R., Govindarajan",
editor="Cohen, Albert",
title="Taming Control Divergence in GPUs through Control Flow Linearization",
booktitle="Compiler Construction",
year="2014",
publisher="Springer Berlin Heidelberg",
address="Berlin, Heidelberg",
pages="133--153",
isbn="978-3-642-54807-9"
}

@misc{arxiv:24:cf_management,
      title={Control Flow Management in Modern GPUs}, 
      author={Mojtaba Abaie Shoushtary and Jordi Tubella Murgadas and Antonio Gonzalez},
      year={2024},
      eprint={2407.02944},
      archivePrefix={arXiv},
      primaryClass={cs.AR},
      url={https://arxiv.org/abs/2407.02944}, 
}

@article{caches:11:characterization,
author = {Wu, Haicheng and Diamos, Gregory and Wang, Jin and Li, Si and Yalamanchili, Sudhakar},
title = {Characterization and transformation of unstructured control flow in bulk synchronous GPU applications},
year = {2012},
issue_date = {May       2012},
publisher = {Sage Publications, Inc.},
address = {USA},
volume = {26},
number = {2},
issn = {1094-3420},
url = {https://doi.org/10.1177/1094342011434814},
doi = {10.1177/1094342011434814},
journal = {Int. J. High Perform. Comput. Appl.},
month = may,
pages = {170–185},
numpages = {16}
}

@INPROCEEDINGS{cgo:22:dram,
  author={Saumya, Charitha and Sundararajah, Kirshanthan and Kulkarni, Milind},
  booktitle={2022 IEEE/ACM International Symposium on Code Generation and Optimization (CGO)}, 
  title={DARM: Control-Flow Melding for SIMT Thread Divergence Reduction}, 
  year={2022},
  volume={},
  number={},
  pages={1-13},
  doi={10.1109/CGO53902.2022.9741285}}

@INPROCEEDINGS{isca:12:capri,
  author={Rhu, Minsoo and Erez, Mattan},
  booktitle={2012 39th Annual International Symposium on Computer Architecture (ISCA)}, 
  title={CAPRI: Prediction of compaction-adequacy for handling control-divergence in GPGPU architectures}, 
  year={2012},
  volume={},
  number={},
  pages={61-71},
  doi={10.1109/ISCA.2012.6237006}}

@INPROCEEDINGS{hpca:13:dual-path,
  author={Rhu, Minsoo and Erez, Mattan},
  booktitle={2013 IEEE 19th International Symposium on High Performance Computer Architecture (HPCA)}, 
  title={The dual-path execution model for efficient GPU control flow}, 
  year={2013},
  volume={},
  number={},
  pages={591-602},
  doi={10.1109/HPCA.2013.6522352}}

@INPROCEEDINGS{hpca:14:multipath,
  author={ElTantawy, Ahmed and Ma, Jessica Wenjie and O'Connor, Mike and Aamodt, Tor M.},
  booktitle={2014 IEEE 20th International Symposium on High Performance Computer Architecture (HPCA)}, 
  title={A scalable multi-path microarchitecture for efficient GPU control flow}, 
  year={2014},
  volume={},
  number={},
  pages={248-259},
  doi={10.1109/HPCA.2014.6835936}}

@INPROCEEDINGS{hpcs:16:cf_restructuring,
  author={Reissmann, Nico and Falch, Thomas L. and Bjørnseth, Benjamin A. and Bahmann, Helge and Christian Meyer, Jan and Jahre, Magnus},
  booktitle={2016 International Conference on High Performance Computing \& Simulation (HPCS)}, 
  title={Efficient control flow restructuring for GPUs}, 
  year={2016},
  volume={},
  number={},
  pages={48-57},
  doi={10.1109/HPCSim.2016.7568315}}

@inproceedings{ics:15:dacache,
author = {Wang, Bin and Yu, Weikuan and Sun, Xian-He and Wang, Xinning},
title = {DaCache: Memory Divergence-Aware GPU Cache Management},
year = {2015},
isbn = {9781450335591},
publisher = {Association for Computing Machinery},
address = {New York, NY, USA},
url = {https://doi.org/10.1145/2751205.2751239},
doi = {10.1145/2751205.2751239},
booktitle = {Proceedings of the 29th ACM on International Conference on Supercomputing},
pages = {89–98},
numpages = {10},
location = {Newport Beach, California, USA},
series = {ICS '15}
}

@INPROCEEDINGS{micro:14:exploringDSE,
  author={Lee, Yunsup and Grover, Vinod and Krashinsky, Ronny and Stephenson, Mark and Keckler, Stephen W. and Asanovic, Krste},
  booktitle={2014 47th Annual IEEE/ACM International Symposium on Microarchitecture}, 
  title={Exploring the Design Space of SPMD Divergence Management on Data-Parallel Architectures}, 
  year={2014},
  volume={},
  number={},
  pages={101-113},
  doi={10.1109/MICRO.2014.48}}

@INPROCEEDINGS{pact:11:divergence,
  author={Coutinho, Bruno and Sampaio, Diogo and Pereira, Fernando Magno Quintao and Meira Jr., Wagner},
  booktitle={2011 International Conference on Parallel Architectures and Compilation Techniques}, 
  title={Divergence Analysis and Optimizations}, 
  year={2011},
  volume={},
  number={},
  pages={320-329},
  doi={10.1109/PACT.2011.63}}

@INPROCEEDINGS{pact:12:data_herding,
  author={Sartori, John and Kumar, Rakesh},
  booktitle={2012 21st International Conference on Parallel Architectures and Compilation Techniques (PACT)}, 
  title={Branch and data herding: Reducing control and memory divergence for error-tolerant GPU applications}, 
  year={2012},
  volume={},
  number={},
  pages={427-428},
  doi={}}

@inproceedings{popl:03:folklore,
author = {Carter, Larry and Ferrante, Jeanne and Thomborson, Clark},
title = {Folklore confirmed: reducible flow graphs are exponentially larger},
year = {2003},
isbn = {1581136285},
publisher = {Association for Computing Machinery},
address = {New York, NY, USA},
url = {https://doi.org/10.1145/604131.604141},
doi = {10.1145/604131.604141},
booktitle = {Proceedings of the 30th ACM SIGPLAN-SIGACT Symposium on Principles of Programming Languages},
pages = {106–114},
numpages = {9},
location = {New Orleans, Louisiana, USA},
series = {POPL '03}
}

@inproceedings{asplos:25:vringo,
author = {Kim, Hansung and Yan, Ruohan Richard and You, Joshua and Yang, Tieliang Vamber and Shao, Yakun Sophia},
title = {Virgo: Cluster-level Matrix Unit Integration in GPUs for Scalability and Energy Efficiency},
year = {2025},
isbn = {9798400710797},
publisher = {Association for Computing Machinery},
address = {New York, NY, USA},
url = {https://doi.org/10.1145/3676641.3716281},
doi = {10.1145/3676641.3716281},
booktitle = {Proceedings of the 30th ACM International Conference on Architectural Support for Programming Languages and Operating Systems, Volume 2},
pages = {1382–1399},
numpages = {18},
location = {Rotterdam, Netherlands},
series = {ASPLOS '25}
}

@misc{dateworkshop:25:pu,
      title={Hardware vs. Software Implementation of Warp-Level Features in Vortex RISC-V GPU}, 
      author={Huanzhi Pu and Rishabh Ravi and Shinnung Jeong and Udit Subramanya and Euijun Chung and Jisheng Zhao and Chihyo Ahn and Hyesoon Kim},
      year={2025},
      eprint={2505.03102},
      archivePrefix={arXiv},
      primaryClass={cs.AR},
      url={https://arxiv.org/abs/2505.03102}, 
}

@misc{dateworkshop:25:Bonet,
      title={A RISC-V Multicore and GPU SoC Platform with a Qualifiable Software Stack for Safety Critical Systems}, 
      author={Marc Solé i Bonet and Jannis Wolf and Leonidas Kosmidis},
      year={2025},
      eprint={2502.21027},
      archivePrefix={arXiv},
      primaryClass={cs.AR},
      url={https://arxiv.org/abs/2502.21027}, 
}

@INPROCEEDINGS{iccd:24:ventus,
  author={Li, Jingzhou and Yang, Kexiang and Jin, Chufeng and Liu, Xudong and Yang, Zexia and Yu, Fangfei and Shi, Yujie and Ma, Mingyuan and Kong, Li and Zhou, Jing and Wu, Hualin and He, Hu},
  booktitle={2024 IEEE 42nd International Conference on Computer Design (ICCD)}, 
  title={Ventus: A High-performance Open-source GPGPU Based on RISC-V and Its Vector Extension}, 
  year={2024},
  volume={},
  number={},
  pages={276-279},
  doi={10.1109/ICCD63220.2024.00049}}

@misc{arxiv:25:egpu,
      title={e-GPU: An Open-Source and Configurable RISC-V Graphic Processing Unit for TinyAI Applications}, 
      author={Simone Machetti and Pasquale Davide Schiavone and Lara Orlandic and Darong Huang and Deniz Kasap and Giovanni Ansaloni and David Atienza},
      year={2025},
      eprint={2505.08421},
      archivePrefix={arXiv},
      primaryClass={cs.AR},
      url={https://arxiv.org/abs/2505.08421}, 
}

@INPROCEEDINGS{fpl:12:guppy,
  author={Al-Dujaili, Abdullah and Deragisch, Florian and Hagiescu, Andrei and Wong, Weng-Fai},
  booktitle={2012 International Conference on Field-Programmable Technology}, 
  title={Guppy: A GPU-like soft-core processor}, 
  year={2012},
  volume={},
  number={},
  pages={57-60},
  doi={10.1109/FPT.2012.6412112}}

@inproceedings{fpga:16:fgpu,
author = {Al Kadi, Muhammed and Janssen, Benedikt and Huebner, Michael},
title = {FGPU: An SIMT-Architecture for FPGAs},
year = {2016},
isbn = {9781450338561},
publisher = {Association for Computing Machinery},
address = {New York, NY, USA},
url = {https://doi.org/10.1145/2847263.2847273},
doi = {10.1145/2847263.2847273},
booktitle = {Proceedings of the 2016 ACM/SIGDA International Symposium on Field-Programmable Gate Arrays},
pages = {254–263},
numpages = {10},
location = {Monterey, California, USA},
series = {FPGA '16}
}

@inproceedings{micro:17:scrach,
  title={SCRATCH: An end-to-end application-aware soft-GPGPU architecture and trimming tool},
  author={Duarte, Pedro and Tomas, Pedro and Falcao, Gabriel},
  booktitle={Proceedings of the 50th Annual IEEE/ACM International Symposium on Microarchitecture},
  pages={165--177},
  year={2017}
}

@INPROCEEDINGS{fpl:21:ma,
  author={Ma, Rui and Hsu, Jia-Ching and Tan, Tian and Nurvitadhi, Eriko and Vivekanandham, Rajesh and Dasu, Aravind and Langhammer, Martin and Chiou, Derek},
  booktitle={2021 31st International Conference on Field-Programmable Logic and Applications (FPL)}, 
  title={DO-GPU: Domain Optimizable Soft GPUs}, 
  year={2021},
  volume={},
  number={},
  pages={140-144},
  doi={10.1109/FPL53798.2021.00031}}

@article{paper:pocl:2015,
title = {pocl: A Performance-Portable OpenCL Implementation},
journal = {International Journal of Parallel Programming },
year = {2015},
author = {Jääskeläinen, Pekkaand and Sánchez de La Lama, Carlos and Schnetter, Erik and  Raiskila, Kalle and   Takala,Jarmo and   Berg, Heikki}
}

@inproceedings{paper:llvm,
author = {Lattner, Chris and Adve, Vikram},
title = {LLVM: A Compilation Framework for Lifelong Program Analysis \& Transformation},
year = {2004},
publisher = {IEEE Computer Society},
pages = {75},
series = {CGO '04}
}

@article{manual:riscv,
  title={The RISC-V instruction set manual, volume I: User-level ISA, version 2.0},
  author={Waterman, Andrew and Lee, Yunsup and Patterson, David A and Asanovic, Krste},
  journal={EECS Department, University of California, Berkeley, Tech. Rep. UCB/EECS-2014-54},
  pages={4},
  year={2014}
}

@INPROCEEDINGS{micro:07:ipdom,
  author={Fung, Wilson W.L. and Sham, Ivan and Yuan, George and Aamodt, Tor M.},
  booktitle={40th Annual IEEE/ACM International Symposium on Microarchitecture (MICRO 2007)}, 
  title={Dynamic Warp Formation and Scheduling for Efficient GPU Control Flow}, 
  year={2007},
  volume={},
  number={},
  pages={407-420},
  doi={10.1109/MICRO.2007.30}}

@INPROCEEDINGS {fccm:25:softcuda,
author = { Ahn, Chihyo and Han, Ruobing and Subramanya, Udit and Zhao, Jisheng and Tine, Blaise and Kim, Hyesoon },
booktitle = { 2025 IEEE 33rd Annual International Symposium on Field-Programmable Custom Computing Machines (FCCM) },
title = {{ SoftCUDA: Running CUDA on Softcore GPU }},
year = {2025},
volume = {},
ISSN = {},
pages = {138-142},
doi = {10.1109/FCCM62733.2025.00049},
url = {https://doi.ieeecomputersociety.org/10.1109/FCCM62733.2025.00049},
publisher = {IEEE Computer Society},
address = {Los Alamitos, CA, USA},
month =May}

@INPROCEEDINGS {hpca:25:tensor_core,
author = { Nada, Abubakr and Sarda, Giuseppe Maria and Lenormand, Erwan },
booktitle = { 2025 IEEE International Symposium on High Performance Computer Architecture (HPCA) },
title = {{ Cooperative Warp Execution in Tensor Core for RISC-V GPGPU }},
year = {2025},
volume = {},
ISSN = {},
pages = {1422-1436},
doi = {10.1109/HPCA61900.2025.00107},
url = {https://doi.ieeecomputersociety.org/10.1109/HPCA61900.2025.00107},
publisher = {IEEE Computer Society},
address = {Los Alamitos, CA, USA},
month =mar}

@INPROCEEDINGS{hpca:25:lmi,
  author={Lee, Jaewon and Chung, Euijun and Singh, Saurabh and Na, Seonjin and Kim, Yonghae and Lee, Jaekyu and Kim, Hyesoon},
  booktitle={2025 IEEE International Symposium on High Performance Computer Architecture (HPCA)}, 
  title={Let-Me-In: (Still) Employing In-pointer Bounds Metadata for Fine-grained GPU Memory Safety}, 
  year={2025},
  volume={},
  number={},
  pages={1648-1661},
  doi={10.1109/HPCA61900.2025.00122}}

@INPROCEEDINGS{carrv:simty,
  author={Sylvain Collange},
  booktitle={First Workshop on Computer Architecture Research with RISC-V}, 
  title={Simty: generalized SIMT execution on RISC-V}, 
  year={2017},
}

@misc{VeriGPU,
  author       = {Hugh Perkins},
  title        = {{VeriGPU}},
  howpublished = {\url{https://github.com/hughperkins/VeriGPU}},
  note         = {Accessed: 2025-07-12}
}

@misc{rv64x,
  author       = {rv64-base},
  title        = {{RV64X}},
  howpublished = {\url{https://github.com/avl-bsuir/rv64x-base}},
  note         = {Accessed: 2025-07-12}
}

@INPROCEEDINGS{iccd:24:SIMTight,
  author={Naylor, Matthew and Joannou, Alexandre and Markettos, A. Theodore and Metzger, Paul and Moore, Simon W. and Jones, Timothy M.},
  booktitle={2024 IEEE 42nd International Conference on Computer Design (ICCD)}, 
  title={Advanced Dynamic Scalarisation for RISC-V GPGPUs}, 
  year={2024},
  volume={},
  number={},
  pages={260-267},
  doi={10.1109/ICCD63220.2024.00047}}

@INPROCEEDINGS{benchmark:rodinia,
  author={Che, Shuai and Boyer, Michael and Meng, Jiayuan and Tarjan, David and Sheaffer, Jeremy W. and Lee, Sang-Ha and Skadron, Kevin},
  booktitle={2009 IEEE International Symposium on Workload Characterization (IISWC)}, 
  title={Rodinia: A benchmark suite for heterogeneous computing}, 
  year={2009},
  volume={},
  number={},
  pages={44-54},
  doi={10.1109/IISWC.2009.5306797}}

@inproceedings{paper:cfg:reducible,
author = {Aho, A. V. and Ullman, J. D.},
title = {Node listings for reducible flow graphs},
year = {1975},
isbn = {9781450374194},
publisher = {Association for Computing Machinery},
address = {New York, NY, USA},
url = {https://doi.org/10.1145/800116.803767},
doi = {10.1145/800116.803767},
booktitle = {Proceedings of the Seventh Annual ACM Symposium on Theory of Computing},
pages = {177–185},
numpages = {9},
location = {Albuquerque, New Mexico, USA},
series = {STOC '75}
}

@inproceedings{paper:cfg:irreducible,
author = {Hecht, Matthew S. and Ullman, Jeffrey D.},
title = {Flow graph reducibility},
year = {1972},
isbn = {9781450374576},
publisher = {Association for Computing Machinery},
address = {New York, NY, USA},
url = {https://doi.org/10.1145/800152.804919},
doi = {10.1145/800152.804919},
booktitle = {Proceedings of the Fourth Annual ACM Symposium on Theory of Computing},
pages = {238–250},
numpages = {13},
location = {Denver, Colorado, USA},
series = {STOC '72}
}

@inproceedings{paper:survey:gpu,
  title = {Impact of {{CUDA}} and {{OpenCL}} on {{Parallel}} and {{Distributed Computing}}},
  booktitle = {2021 8th {{International Conference}} on {{Electrical}} and {{Electronics Engineering}} ({{ICEEE}})},
  author = {Asaduzzaman, Abu and Trent, Alec and Osborne, S. and Aldershof, C. and Sibai, Fadi N.},
  year = {2021},
  month = apr,
  pages = {238--242},
  publisher = {IEEE},
  address = {Antalya, Turkey},
  doi = {10.1109/ICEEE52452.2021.9415927},
  urldate = {2025-01-17},
  copyright = {https://ieeexplore.ieee.org/Xplorehelp/downloads/license-information/IEEE.html},
  isbn = {978-1-66544-071-4},
  langid = {english}
}

@article{paper:survey:parrallelcomputing:14, title={A Survey on Parallel Computing and its Applications in Data-Parallel Problems Using GPU Architectures}, volume={15}, DOI={10.4208/cicp.110113.010813a}, number={2}, journal={Communications in Computational Physics}, author={Navarro, Cristóbal A. and Hitschfeld-Kahler, Nancy and Mateu, Luis}, year={2014}, pages={285–329}}

@INPROCEEDINGS{paper:survey:ai_trends,
  author={Reuther, Albert and Michaleas, Peter and Jones, Michael and Gadepally, Vijay and Samsi, Siddharth and Kepner, Jeremy},
  booktitle={2022 IEEE High Performance Extreme Computing Conference (HPEC)}, 
  title={AI and ML Accelerator Survey and Trends}, 
  year={2022},
  volume={},
  number={},
  pages={1-10},
  doi={10.1109/HPEC55821.2022.9926331}}

@inproceedings{kernel_fission_benchmark_suite,
  title = {A {{Benchmark Suite}} for {{Improving Performance Portability}} of the {{SYCL Programming Model}}},
  booktitle = {2023 {{IEEE International Symposium}} on {{Performance Analysis}} of {{Systems}} and {{Software}} ({{ISPASS}})},
  author = {Jin, Zheming and Vetter, Jeffrey S.},
  date = {2023-04},
  pages = {325--327},
  publisher = {IEEE},
  location = {Raleigh, NC, USA},
  doi = {10.1109/ISPASS57527.2023.00041},
  url = {https://ieeexplore.ieee.org/document/10158214/},
  urldate = {2025-04-26},
  eventtitle = {2023 {{IEEE International Symposium}} on {{Performance Analysis}} of {{Systems}} and {{Software}} ({{ISPASS}})},
  isbn = {9798350397390},
  langid = {english}
}

@misc{benchmark:nvidia_openCL,
  author = {Nvidia},
  title = {Open Computing Language OpenCL},
  url = {https://developer.nvidia.com/opencl}
}

@inproceedings{benchmark:parboil,
  title={Parboil: A Revised Benchmark Suite for Scientific and Commercial Throughput Computing},
  author={John A. Stratton and Christopher I. Rodrigues and I-Jui Sung and Nady Obeid and Li-Wen Chang and Nasser Anssari and Geng Liu and Wen-mei W. Hwu},
  year={2012},
  url={https://api.semanticscholar.org/CorpusID:497928}
}
